\PassOptionsToPackage{table}{xcolor}

\documentclass[acmsmall]{acmart}
\AtBeginDocument{}

\setcopyright{cc}
\acmJournal{PACMNET}
\acmYear{2026} \acmVolume{4} \acmNumber{CoNEXT1} \acmArticle{6}
\acmMonth{3} \acmPrice{} \acmDOI{10.1145/3786289}

\usepackage{array}
\usepackage{makecell}
\usepackage{tikz}
\usepackage{amsmath}
\usepackage{booktabs}
\usepackage{algorithm}
\usepackage[noend]{algpseudocode}
\usepackage{xspace}
\usepackage{multirow} % for \multirow
\usepackage{graphicx}
\usepackage{caption}
\usepackage{forest}
\usepackage{subcaption}
\usepackage{adjustbox}
\usepackage{wrapfig}
\usepackage{colortbl}
\usepackage{etoolbox}
\usepackage{mathtools}
\usepackage{threeparttable} % for tables with notes
\usepackage{cleveref}
\usepackage{color,soul}
\usepackage{diagbox}
\usepackage{enumitem}
\usepackage{subcaption}
\usepackage{caption}
\usepackage{siunitx}
\crefformat{section}{\S#2\color{blue}#1#3} % see manual of cleveref,
\crefformat{subsection}{\S#2\color{blue}#1#3}
\crefformat{subsubsection}{\S#2#1#3}

\begin{document}
\newenvironment{outline}{\color{black!60}}{\par\vspace{\baselineskip}}
\newcommand{\system}{{\textsc{NetSSM}}\xspace}
\newcommand{\systemnormal}{{NetSSM}\xspace}
\newcommand{\sysname}{{\textsc{NetSSM}}\xspace}
\def\ie{{\textit{i.e.,}}~}
\def\eg{{\textit{e.g.,}}~}
\def\cf{{\textit{c.f.,}}~}
\def\etal{{\textit{et al.}~}}

\title{NetSSM: Multi-Flow and State-Aware Network Trace Generation using State Space Models}

\author{Andrew Chu}
\authornote{Equal contribution.}
\affiliation{%
	\institution{University of Chicago}
	\city{Chicago}
	\state{Illinois}
	\country{USA}
}

\author{Xi Jiang}
\authornotemark[1]
\affiliation{%
	\institution{University of Chicago}
	\city{Chicago}
	\state{Illinois}
	\country{USA}
}

\author{Shinan Liu}
\affiliation{%
	\institution{University of Hong Kong}
	\city{Hong Kong}
	\country{Hong Kong}
}

\author{Arjun Bhagoji}
\affiliation{%
	\institution{IIT Bombay}
	\city{Mumbai}
	\country{India}
}

\author{Francesco Bronzino}
\affiliation{%
	\institution{École Normale Supérieure de Lyon; Institut Universitaire de France}
	\city{Lyon}
	\country{France}
}

\author{Paul Schmitt}
\affiliation{%
	\institution{California Polytechnic State University}
	\city{San Luis Obispo}
	\state{California}
	\country{USA}
}

\author{Nick Feamster}
\affiliation{%
	\institution{University of Chicago}
	\city{Chicago}
	\state{Illinois}
	\country{USA}
}
\renewcommand{\shortauthors}{Andrew Chu et al.}

\begin{abstract}

	Access to raw network traffic data is essential for many computer networking tasks, from traffic modeling to performance evaluation.
	Unfortunately, this data is scarce due to high collection costs and governance rules.
	Previous efforts explore this challenge by generating synthetic network data, but fail to reliably handle multi-flow sessions, struggle to reason about stateful communication in moderate to long-duration network sessions, and lack robust evaluations tied to real-world utility. We propose a new method based on state space models called \sysname that generates raw network traffic at the packet-level granularity.
	Our approach captures interactions between multiple, interleaved flows -- an objective unexplored in prior work -- and effectively reasons about flow-state in sessions to capture traffic characteristics.
	\system accomplishes this by training with a context window more than $8\times$ longer, and produces traces up to $78\times$ longer than existing transformer-based raw packet generators.
	Evaluation results show that \system generates high-fidelity traces that outperform prior efforts in existing benchmarks. We also find that \system's traces have high semantic similarity to real network data regarding compliance with standard protocol requirements and flow and session-level traffic characteristics.

\end{abstract}

%%
%% The code below is generated by the tool at http://dl.acm.org/ccs.cfm.
%% Please copy and paste the code instead of the example below.
%%


%%
%% Keywords. The author(s) should pick words that accurately describe
%% the work being presented. Separate the keywords with commas.

% For Articles V4net001-V4net010 use:
\received{June 2025}
\received[revised]{November 2025}
\received[accepted]{December 2025}

\maketitle
\section{Introduction}\label{sec:intro}

There is high demand for representative, scalable network data, driven by applications in security analysis, traffic modeling, and performance evaluation~\cite{singhal2013state, nprint, 5504793, paxson1999bro, baker2004cisco}.
Unfortunately, acquiring large-scale, high-fidelity network data is difficult due to data governance policies, and high collection costs~\cite{abt2014we, de2023survey}.
In response, methods have been developed to generate synthetic network data that accurately replicates real networks, allowing researchers and practitioners to test, evaluate, and model scenarios while minimizing collection overhead and obstacles in accessibility.

Existing methods for generating synthetic network data output this data in two forms: (1) sequences of single or multiple derived network \textit{traffic attributes}, such as flow statistics (\eg duration, average packet size), packet header fields (\eg IP flags, addresses), or metadata (\eg web page views, event types) and (2) raw packet capture (PCAP) traces.
Generators producing traffic attributes can be used to replicate patterns in arbitrarily long network communications.
Generators producing PCAP traces capture the verbose, detailed, communication exchanged between hosts, and commodity packet analyzers (\eg Wireshark) can analyze their resulting PCAPs.

Unfortunately, current methods for either output format have limitations that impact their practical use.
Traffic attribute generators cannot reason about the raw contents of stateful protocols, such as TCP, and require retraining to learn the patterns of new targets in a session.
Raw packet generators are limited in the length of traces they can train on and produce and, thus, may not capture meaningful communication between nodes beyond initial connection setup.
Further, neither generator type can reliably produce data for sessions comprised of more than a single flow, preventing them from being applied to various workloads in the real world, where interleaved, multi-flow communication is common (\eg distributed systems, IoT).
Finally, current methods for evaluating the quality of synthetic network data (\ie statistical similarity to real-world traces and downstream performance of ML models trained on synthetic data) are insufficient.
Synthetic data that perform well in, or towards, these evaluations can still fall short in scenarios that require analysis of multi-flow interactions or stateful behaviors in network traffic (\eg QoE estimation~\cite{sharma2023estimating}, application fingerprinting~\cite{li2022foap}).
Thus, determining the criteria for what qualities or characteristics make synthetic network data ``good'' is an ongoing area of research.

In this paper, we present \system, a raw packet generator for network traffic data built on the recently proposed structured selective state space model (Mamba) architecture. \system bridges the gap between traffic attribute and raw packet generators by combining the former's length-scaling capabilities with the latter's comprehensive packet-level detail. This enables \system to capture a substantially wider range of target events while retaining the ability to capture inter- and intra-packet dependencies across any protocol and layer.
Furthermore, the sequential, stateful nature of how \system learns network data allows it to generate sessions comprised of multiple interleaved flows with high fidelity, addressing a limitation of existing methods.

We evaluate \system on social media, video conferencing, and video streaming traffic.
First, we assess its performance using established metrics of synthetic network data fidelity (statistical similarity and downstream ML performance).
We then evaluate \system's \textit{semantic similarity}, testing how well its generated data aligns with the behavioral characteristics of real-world network communication. Finally, we verify that \system's traces are both protocol compliant, and mimic, rather than memorize patterns in training data.
This analysis aims to offer a more functional and application-oriented perspective on the quality of synthetic data, emphasizing its practical utility beyond statistical resemblance.
Our main
contributions are:

\begin{itemize}[noitemsep,topsep=0pt]

	\item \textbf{Synthetic multi-flow sessions.} \system's recurrent nature
	      enables it to produce traces for sessions comprised of both single flow
	      and multi-interleaved flows, with high fidelity.
	      Multi-flow trace generation is a new contribution largely unexplored in prior generators.

	\item \textbf{Capturing flow-state-dependent session events.} \system trains
	      using a context window more than $8\times$ longer, and produces traces up
	      to $78\times$ longer than existing transformer-based raw packet
	      generators.
	      This enables it to learn from and output traces that capture flow-state-dependent events occurring later in a session that rely on early connection setup, or multiple interactions between flows and/or packets.

	\item \textbf{Superior performance on existing benchmarks.} \sysname
	      outperforms current state-of-the-art network data generators in existing
	      benchmarks.
	      In statistical similarity, \system achieves an average Jensen-Shannon Divergence across generated traffic attributes of $0.05$, versus $0.18$ and $0.06$ for NetShare~\cite{yin2022practical} and NetDiffusion~\cite{jiang2024netdiffusion}, respectively.
	      In performance of downstream ML models trained on synthetic data, a random forest classifier trained entirely on synthetic \system data achieves accuracy of $0.97$ on held-out ground truth data, compared to $0.13$ and $0.16$ for NetShare and NetDiffusion respectively.

	\item \textbf{Behaviorally accurate and protocol-adherent traffic.} \sysname
	      generates synthetic traffic with high semantic similarity to real traces.
	      This traffic can (1) capture application-specific traffic patterns, and (2) show robust session-level compliance with standard TCP protocol requirements, capturing both correct stateful behavior and common real-world anomalies (\eg partial teardowns, conflicting flags).
	      For (1), \system can generate traces that capture the sequential communication and distributional patterns in traffic, even presented with complex, multi-flow traffic comprised of multiple steps (\eg setup with CDN endpoints before video segment downloads for video streaming traffic).
\end{itemize}

\system's code and training datasets are open sourced at \href{https://github.com/noise-lab/netssm}{https://github.com/noise-lab/netssm}.

\section{Related Work}\label{sec:related}

Techniques for generating synthetic network data aim to replicate the characteristics of real-world communication between networked devices, either through higher-level traffic attributes about packets or a session, or raw packet captures.
Traffic generators can be categorized into two main approaches: traffic attribute generators and raw packet generators.

\subsection{Traffic Attribute Generators}

Traffic attribute generators use simulation or machine learning to produce higher-level data describing networked communication.
Simulation-based approaches were the earliest method for synthesizing network data, using user-defined templates to configure a simulated network (\eg topology, link specifications, workload), and replaying or emulating communication on this network to produce traffic attributes relevant to the simulated network.
Notable efforts in this approach include NS-3~\cite{henderson2008network}, TRex~\cite{ciscotrex2023}, and others \cite{botta2012tool,buhler2022generating,lacage2006yet}, which remain popular due to their configurability, versatility, and relative inexpensiveness.
Unfortunately, these methods' simulated traffic may not model the variability and unpredictability inherent in actual network conditions~\cite{campanile2020computer,swann2021tools}, and thus may fail to capture the nuances of real-world traffic exchange.

Machine learning-based approaches adopt techniques for time-series forecasting to learn signals in a given continuous stream of input.
These models isolate the fine-grained variations in one or more traffic attributes and produce data statistically similar to real-world traffic, further reinforced by offering improved performance when used in downstream ML-based tasks (\eg service recognition, anomaly detection).
Additionally, this data can have arbitrary length, as the generating model learns from only a single or small set of continuous traffic attribute values.
Early ML-based traffic attribute generators include Lin \etal's DoppelGANger which uses a generative adversarial network (GAN) to produce sequences of single traffic attribute values~\cite{lin2020using}, and Yin \etal's NetShare which builds on DoppelGANger to output more expressive sets of aggregate traffic attributes (\eg duration, packet count), or more comprehensive sets of packet-level header field values (\eg time-to-live [TTL], protocol flags)~\cite{yin2022practical}.
Zhang \etal's NetDiff uses both diffusion and transformers to try to better encode patterns in traffic attributes and use this encoding to better inform generation, specifically for mobile network data~\cite{zhang2024netdiff}.
One limitation of these models is that when modeling raw packet contents, they only support learning and generating values from Layer 3 and below (plus transport-layer port numbers) in the OSI model.
Thus, they cannot model interactions or attributes in stateful protocols (\eg TCP).

\subsection{Raw Packet Generators}

Raw packet generators use simulation or machine learning to output synthetic network traffic in the form of verbose, raw PCAPs.
The same simulation-based approaches used to produce traffic attributes can be used to produce raw traces, where the simulated communication between nodes is collected (versus summarized to yield traffic attributes) and written to a trace.
Unfortunately, the same shortcoming in expressiveness also exists for these simulators for this output granularity.

Machine learning-based approaches train on raw packet data and generate the byte-level values that comprise the packets of a session.
Whereas traffic attribute generators are designed to learn from and capture variations in values over time implicit in a given sequence, raw packet generators learn from and capture the inter- and intra-packet relationships contained in a trace's raw contents, from which traffic attributes can be extracted.
Operating at the packet level, these generators can model protocols at any layer.
Evaluated under the same metrics, raw packet generators have comparable or better statistical similarity and downstream ML-task performance than traffic attribute generators, and their verbose PCAP format may be more versatile for later analysis and feature extraction. For example, Jiang \etal's NetDiffusion uses a text-to-image diffusion model trained on image representations of network traces to generate images with specific traffic characteristics and are convertible back to PCAP form~\cite{jiang2024netdiffusion}. Qu \etal's TrafficGPT is a transformer decoder that trains on, and produces tokens corresponding to raw bytes of PCAP traces~\cite{qu2024trafficgpt}.
A key drawback to existing diffusion and transformer-based raw packet generators is their relatively short limit in training context and output length (\ie learning from and producing PCAPs with maximum lengths of $1{,}024$ packets for NetDiffusion and $113-128$\footnote{Using packet lengths of $94$/$106$ tokens from our evaluation case studies, for TrafficGPT's max generation length of $12{,}032$ tokens.} packets for TrafficGPT), which may fail to capture target events in exchanged communication.
Most recently, Chu and Jiang~\etal proposed using SSMs, specifically Mamba-1, to generate synthetic traces~\cite{chu2024feasibility}.
Our work improves on this effort by using the Mamba-2 architecture (allowing for larger modeled state and faster training), training on longer contexts ($100{,}000$ versus $50{,}000$ tokens), producing multi-flow traces (versus only single-flow) and presenting more detailed evaluation of generation quality.

\section{State Space Models for Network Traffic Generation}
\label{sec:ssm_background}

Much communication between networked devices is stateful, and these exchanges may span long sequences of packets for multiple steps (\eg setup, payload download, teardown). Our choice of Mamba~\cite{gu2023mamba,dao2024mamba2}, a line of selective structured SSMs, accommodates these characteristics.
In this section, we provide background on SSMs, specifically, the Mamba model (Section~\ref{subsec:ssm_technical}), and compare Mamba against the existing approaches in raw packet generators (Section~\ref{subsec:ssm_comparison}).

\subsection{Background: State Space Models and Mamba}\label{subsec:ssm_technical}

SSMs are probabilistic graphical models built on the control engineering concept of a state space~\cite{kalman1960new}.
Similar to Hidden Markov Models, SSMs model discrete observations over time, but use continuous, as opposed to discrete, latent variables.
SSMs encode a running hidden state representative of prior observed context of input using recurrent scans, and use this state to calculate an output for a given unobserved input.
Specifically, SSMs use first-order ordinary linear differential equations to capture the relationship (output) between unobserved variables (state) and a series of continuous observations (input), irrespective of time (\ie is linear time-invariant [LTI]).
Unfortunately, SSMs suffer from the same pitfall as other recurrently updating models, in that over time, information about data earlier in an input becomes increasingly compressed in the hidden state.
This leads to the ``vanishing gradient,'' where the model can no longer recall dependencies between inputs.
Prior works by Gu~\etal and Voelker~\etal remedy this challenge by fixing the state matrix used in SSMs, resulting in improved model performance for recalling long-range dependencies~\cite{gu2020hippo,voelker2019legendre}.
Follow-up works by Gu~\etal provide additional improvements to the SSM, improving training efficiency for practical use via convolutional kernel (S4 \cite{gu2021s4}) and sequence modeling performance via a selection mechanism and a fixed state matrix (Mamba, Mamba-2 \cite{gu2023mamba,dao2024mamba2}).

Specifically, Mamba builds on S4, and additionally implements two modifications to the general SSM that provide \textit{structure} and \textit{selection}.
It implements structure by replacing the general SSM state matrix (typically randomly initialized) with a HiPPO matrix~\cite{gu2020hippo}, which enforces a probability measure for dictating how the SSM state is compressed.
This, in effect, remedies the vanishing gradient and improves the Mamba SSM's ability to model long-range dependencies in sequences.
For selection, the general LTI SSM lacks expressiveness, \ie all discrete inputs compressed in the state affect the state with equal weighting.
In language modeling, this prevents semantically important ``keywords'' from more heavily influencing the SSM state and developing a better understanding of input.
Mamba improves expressiveness by removing the LTI quality of the general SSM and makes the model \textit{time-variant}, in which the state is calculated using learned (rather than fixed) functions of the inputs.
Mamba's structure and selection modifications to the general SSM architecture provide competitive performance against conventional transformer-based approaches for sequence modeling, with better scaling (linear versus quadratic).

\subsection{Why Mamba?} \label{subsec:ssm_comparison}

We select the Mamba architecture because it is inherently suited to the nature of network data, specifically the stateful nature of a large portion of network traffic (\eg TCP flows).
Communication between hosts often explicitly depends on the sequential exchange of packets to ensure correct data assembly and to maintain the connection.
This can be mapped to the recurrent quality of the \textit{state space} architecture, where the model sequentially updates the hidden state on each new input.
In our use of Mamba for synthetic trace generation, this enables \system to effectively learn from and produce sessions composed of multiple flows.
In contrast, prior traffic attribute and raw packet generators can only operate within the scope of single-flow sessions.

The architecture's convolutional kernel further complements the network domain by enabling updates to be performed in parallel, allowing the model to train on substantially long communication while still implicitly capturing sequential dependencies.
As such, Mamba is a much more ``natural'' fit for modeling network data compared to prior raw packet generators.
Diffusion-based approaches require abstracting network data to a different domain (\ie images in NetDiffusion), and further generate traces based on signals from the entire trace, neglecting the sequential delivery of network traffic.
Transformer-based models likewise learn input semantics in a completely parallel fashion, where attention is calculated per token of a sequence, against all other tokens in the sequence simultaneously, also not strictly sequentially.
The completely parallel computation nature of either approach is also resource-intensive. \system can generate traces roughly $10\times$ longer than NetDiffusion and $78\times$ longer than TrafficGPT. This is a key improvement, allowing \system to capture flow-state-dependent sessions events that manifest only after substantial setup has occurred, and thus may not be captured with other models.
For instance, in video streaming traffic (generated/evaluated in Section~\ref{subsubsec:semantic_sim_segments}), a flow's representative state for downloading audio/video segments may not be reached until after a few hundred or thousand packets.

\section{\systemnormal}\label{sec:netmamba}

\begin{figure}[t]
  \centering
  \includegraphics[width=\linewidth]{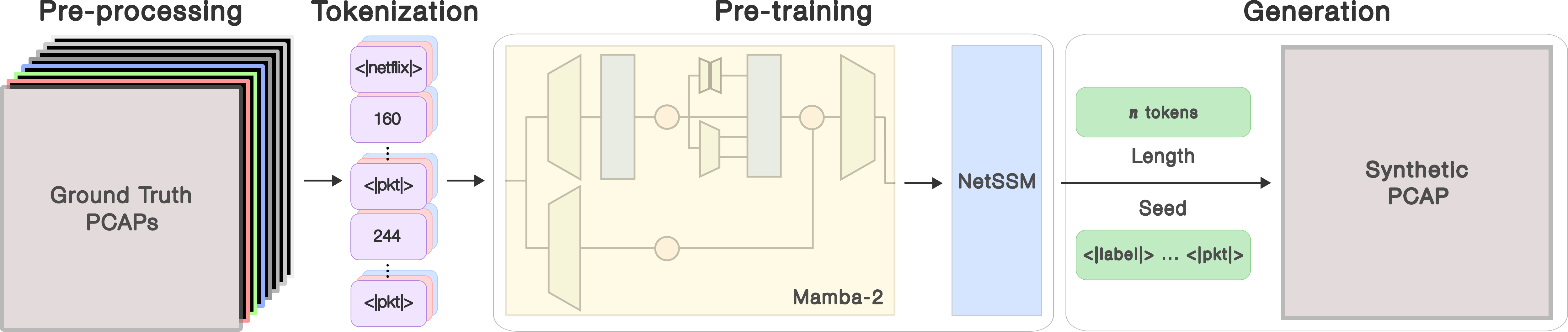}
  \caption{Overview of the \system pipeline.}
  \label{fig:pipeline}
  \Description[Model training and generation process.]{Model training and generation process.}
\end{figure}

Motivated by shortcomings in existing synthetic network data generators, and strong alignment with the operation and capabilities of SSMs and the qualities of networked communications, we present \system, a new raw packet generator.
\system uses the Mamba-2-backbone, and is trained the raw byte contents of packets, to synthesize packet traces.
Figure~\ref{fig:pipeline} provides an overview of the \system pipeline, and we provide details for each pipeline step below.

\subsection{Pipeline Overview}
\subsubsection{Pre-processing Networking Data.}\label{subsec:netmamba_preproc}

Input to \system are sequences of the raw bytes which comprise the packets in a session trace.
Specifically, \system parses the Packet Data field of each packet record in a PCAP file~\cite{ietf-opsawg-pcap-06} to a representative format that aligns with the token-based, sequence generation objective of the Mamba SSM.

\noindent\textbf{Tokenization.}\label{subsubsec:tok}
We define a custom tokenizer using Huggingface Tokenizers \cite{Moi_HuggingFace_s_Tokenizers_2023} that one-to-one maps the decimal values of the raw bytes comprising each packet to a corresponding token ID in range $[0, 255]$. In this way, \system reasons about the raw contents of networking traffic close to its original form.
This differs from prior work where network data is represented/tokenized at the flow level~\cite{Lin_2022}, as a mix of packet-level and flow attributes \cite{qu2024trafficgpt}, or created using a tokenization algorithm that may map raw bytes to tokens using logic suited to a different domain (\ie WordPiece from NLP)~\cite{meng2023netgpt}.
Our tokenizer also defines label special tokens (\eg \texttt{<|facebook|>} \texttt{<|meet|>}, \texttt{<|netflix|>}) and a packet special token (\texttt{<|pkt|>}) to allow \system to differentiate between traffic dynamics of different workloads, and packet boundaries in sessions.

\noindent\textbf{Creating training data.}\label{subsubsec:parsing}
We extract input to \system from labeled (\ie the workload/service type of collected traffic is known) collections of PCAPs based on the desired modeling granularity, \ie single-flow or multi-flow sessions. For single-flow sessions, we use \texttt{pcap-splitter}~\cite{pcapsplitter} to first split the original PCAP into multiple PCAPs, each corresponding to a single comprising flow based on connection (\ie \textit{five-tuple}: source IP, source port, destination IP, destination port, IP protocol).
No pre-processing is needed for multi-flow sessions (\ie captures containing multiple connections/five-tuples).
We then use our custom PCAP parser written in Go, which performs the following tasks: (1) converts the raw bytes comprising each packet in a PCAP to a sequence of 8-bit decimal values (\ie $\text{value} \in [0, 255]$) in string form, (2) delimits each string form packet with \texttt{<|pkt|>} special tokens, and (3) prepends the PCAP's corresponding label special token to the string. Finally, we use the custom \system tokenizer to tokenize the parsed, string-based PCAP data to a format consumable by \system, producing one input sample for each PCAP in a dataset.
Our parser currently supports processing both IPv4 and IPv6 packets, the TCP and UDP transport protocols, and the DNS application protocol.
The parser can easily be extended to additional protocols or workloads by simply adding a new corresponding processing function.

\subsubsection{Pre-training \systemnormal}\label{subsec:nm_models}
Training data are fed into \system to learn the semantics of packets, and correspondingly, flows and sessions.
Specifically, \system treats generating network traffic data as a self-supervised sequence generation problem. During training, \system minimizes the cross-entropy loss function, which measures how well the predicted probabilities for a token at a specific index match the correct token. Because this learning objective is irrespective of the input used, \system is easily extensible to learning the semantics of any protocol or workload. For our experiments with \system, we train using a batch size of one, which allows each input/training sample to be $100{,}000$ tokens in length (the maximum length supported for our experiment setup).
This maximizes the length of packet sequences our model learns from (\ie context length), with $100{,}000$ tokens corresponding to a context of at least $943$ packets (when using different packet representations from our various case studies).

\subsubsection{Generating Synthetic Traces}\label{subsubsec:generation}
Trace generation requires two arguments: a generation seed and length.
The generation seed matches the format of \system's training samples -- a label special token followed by a sequence of any number of full or partial packets represented by their raw-byte contents in decimal form (\eg \texttt{<|amazon|> 188 34 203.
	.. <|pkt|>}).
The seed is used to ``prompt'' \system for generation, equivalent to the ``start token'' or string in NLP generative models.
The generation length dictates the length of output (in tokens, or optionally, packets) that \system should generate.
\system's packet model encodes the generation seed to its latent representation before passing it first through the actual SSM linear system, and second the softmax function.
This results in a set of probabilities each token (\ie byte value) has of being selected as the generation candidate. On each generation step, the single highest probability token is both output by \system and used to update the existing latent representation, becoming the input for the next round of generation.
This procedure continues until the given generation length is satisfied. \system then constructs the intermediate synthetic trace, concatenating the sequence of generated packets represented by their tokenized raw bytes in decimal format, and prepending the label special token.
The pipeline concludes with a simple script which converts the token-based trace representation to a complete PCAP binary, with the option to assign packet inter-arrival times (IATs) to packets by sampling from the IAT distribution of a given ground truth capture (presumably from the same traffic class or workload).

\subsection{What \systemnormal Does and Does Not Do}
\system generates traces of raw packet communication in the form of PCAPs.
These traces can be of arbitrary, user-specified length, and may be comprised of either a single flow (\ie two endpoints) or multiple flows (\ie more than two endpoints).

These traces capture the sequential characteristics of packet communication (and thus, may act as a weak proxy for time). Unfortunately, \system does not extract or parse, and hence, does not learn from and autoregressively generate, the timestamp/IAT values for each packet. We provide additional discussion on this shortcoming and future directions for addressing it in Section~\ref{sec:discussion}.

\section{Evaluation}\label{sec:evaluation}

We evaluate the quality of synthetic data produced by \system through five analyzes: (1) \textit{statistical similarity} between generated and real traffic, (2) \textit{downstream utility} of generated data towards training and improving ML-for-networking models, (3) \textit{semantic similarity} between generated and real traffic, (4) \textit{protocol compliance} between generated and real traffic, and (4) \textit{analysis of memorization} in synthetic \system traffic.
Previous traffic attribute and raw packet generators are measured using metrics of statistical similarity and downstream performance.
We introduce semantic similarity and protocol compliance as additional aspects that should be considered when evaluating synthetic network data models or systems. Finally, we perform analysis on \system's output and show that it is learning to mimic, not memorize, patterns in network data used during training.

\subsection{Statistical Similarity}\label{subsec:statistical}

We first evaluate \system with conventional metrics of statistical similarity used to evaluate prior traffic generators, which assess the byte-wise matching between generated synthetic traces and the ground truth training traces.
Specifically, we train a \sysname model on \textit{single-flow} traces (\ie comprised of a single connection/five-tuple) collected from various types of multimedia traffic.
We examine the single-flow granularity so that we can provide direct comparison against prior work, which all evaluate at this level.
After training, we generate synthetic traces and compare them to their ground truth counterparts. We find that \system's synthetic traces exhibit high statistical similarity to real data at the content level (byte-wise comparisons), outperforming previous synthetic network trace generation methods in various statistical metrics.

\subsubsection{Setup.}\label{subsubsec:sfe_setup}

We evaluate the statistical similarity of traces produced by (1) a base \system model that trains on and produces continuous sessions, and (2) a fine-tuned version of the base model that generates packets comprising distinct flow stages (\eg TCP teardown, characterized by \texttt{ACK} and \texttt{FIN} packets, or data transmission, characterized by \texttt{PUSH} and \texttt{ACK} packets).
Here, we wish to examine if additional fine-tuning can yield performance improvements, particularly in generating these distinct components of networked communication.
Fine-tuned models could be especially useful for applications requiring only subsets of a trace to study key network behaviors (\eg session termination indicators).
We detail the setup for either model below.

%!TEX root = ../paper.tex

\begin{table*}[t]
  \centering
  \caption{Overview of datasets used to train and evaluate \system.}
  \vspace{-3mm}
  \begin{adjustbox}{max width=\textwidth}
    \begin{threeparttable}
      \begin{tabular}{p{2.8cm}>{\raggedleft\arraybackslash}p{3.25cm}>{\raggedleft\arraybackslash}p{1.77cm}>{\raggedleft\arraybackslash}p{2.9cm}>{\centering\arraybackslash}p{2.25cm}>{\raggedleft\arraybackslash}p{1.6cm}>{\raggedleft\arraybackslash}p{1.66cm}}
        % \begin{tabular}{lrrrcrr}
        \toprule
        \multirow{2.5}{*}{\sc{Dataset}} & \multicolumn{1}{c}{\multirow{2.5}{*}{\sc{Source}}} & \multicolumn{1}{c}{\multirow{2.5}{*}{\sc{Evaluation}}} & \multicolumn{2}{c}{\sc{Content Type}} & \multicolumn{2}{c}{\sc{Size}}
        \\
        \cmidrule(lr){4-5} \cmidrule(l){6-7}
        &
        &
        & \sc{Classification}                                    &
        \sc{\# Sub-classes}
        &
        \multicolumn{1}{c}{\sc{Raw}}         &
        \multicolumn{1}{c}{\sc{\# Captures}}
        \\
        \midrule
        \multirow{3}{*}{Multimedia Traffic}  &
        \multirow{1}{*}{Bronzino \etal~
        \cite{bronzino2019inferring}}        &
        \S\ref{subsec:statistical}, \ref{subsec:ml_task},
        \ref{subsec:protocol_compliance}     & \multirow{1}{*}{Video
        Streaming}                           & \multirow{1}{*}{4}
        & \multirow{1}{*}{$6.36$ GiB}
        & \multirow{1}{*}{$273$}
        \\
        & \multirow{1}{*}{MacMillan
        \etal~\cite{macmillan2021measuring}} &
        \S\ref{subsec:statistical}, \ref{subsec:ml_task},
        \ref{subsec:protocol_compliance}     & \multirow{1}{*}{Video
        Conferencing}                        & \multirow{1}{*}{3}
        & \multirow{1}{*}{$17.36$ GiB}
        & \multirow{1}{*}{$339$}
        \\
        & \multirow{1}{*}{Jiang \etal~\cite{jiang2025jiti}}
        & \S\ref{subsec:statistical}, \ref{subsec:ml_task},
        \ref{subsec:protocol_compliance}     & \multirow{1}{*}{Social
        Media}                               & \multirow{1}{*}{3}
        &
        \multirow{1}{*}{$5.40$ GiB}          & \multirow{1}{*}{$151$}
        \\
        % & & & & & \\
        %
        \midrule
        Netflix Streaming                    &
        Bronzino~\etal~\cite{bronzino2019inferring}
        & \S\ref{subsec:semantic_sim},
        \ref{subsec:protocol_compliance}     & Video
        Streaming                            & 1
        & $216.36$ GiB                                           &
        $5{,}882$
        \\
        \midrule
        YouTube Streaming                    &
        Gutterman~\etal~\cite{gutterman2019requet}
        & \S\ref{subsec:semantic_sim}
        & Video
        Streaming                            & 1
        & $2.06$ GiB                                             & $619$
        \\
        \bottomrule
      \end{tabular}
    \end{threeparttable}
  \end{adjustbox}
  \label{tab:dataset_overview}
\end{table*}

\noindent\textbf{Base model.}
We train our \system model for single-flow trace generation using the Multimedia Traffic dataset outlined in Table~\ref{tab:dataset_overview}.
We first pre-process the data using \texttt{pcap-splitter}~\cite{pcapsplitter}, splitting PCAPs into their comprising single-flow PCAPs based on five-tuple, and parse them into the string representations of their raw bytes in decimal form, as described in Section~\ref{subsubsec:parsing}.
We fix each packet to be represented by $94$ tokens, corresponding to the maximum practical lengths of the Ethernet ($14$ bytes), IPv4 ($20$ bytes excluding options), and TCP headers ($60$ bytes including extensions). We train the \system model for this evaluation on TCP traffic only, and do not consider TCP payload, as this data is becoming increasingly encrypted~\cite{rfc8484,rfc7858,letsencrypthttps} and would be noise our model would not learn from.
Next, we create a custom tokenizer following the configuration described in Section~\ref{subsubsec:tok}, defining $10$ label special tokens corresponding to the $10$ distinct applications in our dataset.
We then tokenize all string representations resulting from splitting our data to their single-flows resulting in a final dataset of $27{,}839$ samples.
Finally, we pre-train the single-flow packet \system model on the created dataset using a single NVIDIA A40 48GB GPU for $30$ epochs with a gradient clip value of $1.0$ and default AdamW optimizer with learning rate of $\num{5e-4}$.
We use the same configuration as the smallest publicly available $130$ million parameter pre-trained Mamba-2 (dimension of $768$, $24$ layers), but instead use our custom tokenizer.
We generate traces using the process detailed in Section~\ref{subsubsec:generation}, producing a corresponding synthetic trace for each real trace used in training.
Specifically, we use the first packet from the real training trace in tokenized form, along with its corresponding label as the seed.
We set the generation length as the number of tokens needed to represent the total packets in a corresponding real trace.
This ensures the generated trace contains the same number of packets as the real trace, providing a consistent basis for evaluating the synthetic trace's statistical similarity to the real data.

\noindent\textbf{Fine-tuned model.}
We train the fine-tuned \system model by first creating sub-datasets from the original dataset described above that isolate the packets relevant to specific stages of a flow's lifetime.
These sub-datasets focus on distinct phases of network communication (\eg session initiation, data exchange, session termination).
We use these phase-specific data to fine-tune the base $30$-epoch single-flow \system, using the same next-token prediction objective as the original model but with phase-specific packets as input.
This allows the model to capture the intra-packet and flow dynamics unique to each phase, leading to improvements in both the quality and flexibility of output.
When generating data with the fine-tuned models, we chain outputs from one phase-specific model to the next.
Specifically, the final packet produced by the handshake model serves as the seed for the subsequent data transmission model, while the final packet generated by the data transmission model acts as the seed for the subsequent session teardown model.

\noindent\textbf{Baselines.}
We compare the two \system variants against three prior works: NetDiffusion, TrafficGPT, and NetShare.
We also evaluate against two random generations of flow statistics (uniformly sampled random values across valid attribute ranges, \eg IP addresses/ports, and protocols) and raw packets (random assignment of 1, 0, -1 to indices in the nPrint packet format~\cite{nprint}) to serve as benchmarks for poor fidelity.
Specifically, we train new NetShare and NetDiffusion models on our Multimedia Traffic dataset (for TrafficGPT, we rely on the paper's reported results as it is closed source), and use these models to generate corresponding synthetic traffic attributes, or raw PCAPs, based on each capture in the ground truth dataset.

\subsubsection{Results.}\label{subsubsec:sfe_content_results}

\begin{table}
  \rowcolors{5}{}{lightgray!35}

  \caption{\textbf{Byte-wise statistical similarity.}
  Across generators and data granularities, \sysname traces are most statistically similar to real traffic (divergence/distance metrics $\geq2\times$ lower versus the next best method).}

  \label{tab:statistical_similarity_content}
    \vspace{-3mm}
    \centering \resizebox{\linewidth}{!}{
    \begin{threeparttable}
      \begin{tabular}{l|cccccc|ccc}
        \toprule
        \multicolumn{1}{l}{\multirow{2.25}{*}{\sc{Generation Method}}} &
        & \multicolumn{5}{l}{\sc{Traffic Attribute-level (Avg. JSD)}
        $\downarrow$}                                                  &
        \multicolumn{3}{c}{\sc{Header-level} $\downarrow$}
        \\
        \cmidrule[0.5pt](rl){2-7} \cmidrule[0.5pt](rl){8-10}
        \multicolumn{1}{l}{~}                                          &
        \multicolumn{1}{c}{\sc{SA}}
        & \sc{DA}                                                    &
        \sc{SP}
        & \sc{DP}
        & \sc{PR}                                                    &
        \multicolumn{1}{c}{\sc{Avg.}}
        & \multicolumn{1}{c}{\sc{Avg. JSD}}
        & \sc{Avg. TVD}                                              &
        \sc{Avg. HD}
        \\
        \midrule
        Random Generation (flow statistics)
        & 0.71
        & 0.71                                                       & 0.63
        &
        0.63
        & 0.47
        & 0.63                                                       &
        ---
        & ---
        & ---
        \\
        Random Generation (raw packets)
        & ---
        & ---                                                        & ---
        & ---                                                        &
        ---
        & ---                                                        &
        0.82
        & 0.99
        & 0.95
        \\
        NetShare
        & 0.14
        & 0.19                                                       & 0.29
        & 0.25                                                       & 0.04
        & 0.18                                                       & ---
        & ---
        & ---                                                               \\
        TrafficGPT\tnote{*}
        & 0.13
        & 0.16                                                       & 0.17
        & 0.23
        & ---                                                        & 0.17
        & ---
        & ---                                                        & ---
        \\
        NetDiffusion$^\dagger$
        & 0.00
        & 0.00                                                       & 0.14
        & 0.17                                                       & 0.06
        & 0.06                                                       & 0.04
        & 0.04
        & 0.05                                                              \\
        \sysname (base)
        & 0.12
        & 0.11                                                       & 0.10
        & 0.11                                                       & 0.00
        & 0.09                                                       & 0.02
        & 0.02
        & 0.02
        \\
        \sysname (fine-tuned)
        & 0.06
        & 0.05                                                       &
        0.05
        & 0.06
        & 0.01                                                       & 0.05
        & 0.02                                                       & 0.01
        & 0.02
        \\
        %  \sysname (\chase{on new dataset}) & 0.02 & 0.01 & 0.07 &
        % 0.07 & 0.00 & 0.04 &
        % 0.05 & 0.04 & 0.05 \\
        \bottomrule
      \end{tabular}
      \begin{tablenotes}
        \footnotesize
      \item[]*As reported
        in~\cite{qu2024trafficgpt}.\hfill$^\dagger$Post-generation
        correction applied.
      \end{tablenotes}
    \end{threeparttable}
  }
\end{table}

%New YouTube data:
%Average statistical results across non-overfitting fields:
%
%Jensen-Shannon Divergence: 0.049023
%Total Variation Distance:  0.036122
%Hellinger Distance:        0.053425
%
%
%============================================================
%Summary of Network Feature JSD
%============================================================
%Source IP Address        : 0.023998
%Destination IP Address   : 0.012294
%Source Port              : 0.073788
%Destination Port         : 0.068561
%Protocol                 : 0.004194
%============================================================
%Average JSD (5 features) : 0.036567

We evaluate \sysname's generation fidelity by analyzing the statistical similarity between its synthetic data, and each of these synthetic data's real, ground truth counterpart, consistent with prior evaluations ~\cite{jiang2024netdiffusion, lin2020using, yin2022practical, qu2024trafficgpt}.
Specifically, we calculate three distributional distance metrics: Jensen-Shannon Divergence (JSD), Total Variation Distance (TVD), and Hellinger Distance (HD), where lower values indicate closer alignment to the ground truth.
We calculate these metrics at two levels: (1) traffic attribute-level for direct comparison against all prior works, and (2) header-level for comparison against NetDiffusion.
We compute traffic-attribute level metrics for the fields of source and destination IP addresses and ports (SA, DA, SP, DP) and IP protocol (PR). Table~\ref{tab:statistical_similarity_content} presents the overview of statistical similarity for each generator and level.

\noindent\textbf{Traffic attribute-level similarity.}
We extract the ground truth traffic attributes from the ground truth traces, and compute the distance metrics between these values and their synthetic counterparts.
Specifically, for \system and NetDiffusion, we extract traffic attributes from each generated PCAP corresponding to a ground truth capture.
NetShare directly generates traffic attributes, and thus does not require additional parsing.
For brevity, Table~\ref{tab:statistical_similarity_content} shows only the average JSD for each field, though the TVD and HD are highly similar ($\pm 0.02$). Between the base and fine-tuned variants, \system achieves the best or second-best average JSD in all fields.

\noindent\textbf{Header-level similarity.}
We compare statistical similarity at the header level by calculating the three distance metrics across each bit position in the nPrint~\cite{nprint} representation of a trace, for all packets' TCP headers.
We exclude NetShare (only generates a subset of header fields) and TrafficGPT (closed source preventing further analysis) from this evaluation.
\sysname consistently outperforms NetDiffusion in all metrics with distances as low as $0{.}01$ in the fine-tuned \system variant.
While the distance delta between NetDiffusion may appear only marginal, we re-emphasize that the comparison is performed \textit{after} post-generation correction in NetDiffusion is applied.
To illustrate, some NetDiffusion-generated traces in this experiment were unparseable by packet analysis tools prior to applying the heuristic-based fix. In contrast, \system requires no post-generation correction and yields better statistical similarity.

\subsection{Downstream Utility}\label{subsec:ml_task}

We examine the performance of downstream ML models trained on synthetic network data to evaluate this data's quality in practical application.
Specifically, we train two types of classifiers that focus on (1) application-level classification (\eg YouTube, Amazon) and (2) service-type-level classification (\eg video streaming, web browsing).

\subsubsection{Setup.}\label{subsubsec:downstream_setup}

To test the utility of synthetic data in augmenting downstream model training, we create \textit{downstream training datasets} comprised of packet header-level features determined via nPrintML~\cite{holland2021new}.
We extract these features from both the ground truth Multimedia Traffic dataset, and three sets of synthetic data generated by \system, NetDiffusion, and NetShare models trained on this dataset.
Each downstream training dataset uses different \textit{mixing rates} that represent the proportion of synthetic data used to replace original real data in the dataset.
We create a new dataset at each $10\%$ inclusive increments, resulting in $33$ downstream training datasets ($11$ for each model).
For example, a downstream training dataset with a 20\% mixing rate contains $80\%$ real data, and $20\%$ synthetic data.
We train three different types of ML classifiers (Decision Trees [DT], Random Forest [RF], and Support Vector Machines [SVM]) on these downstream datasets, resulting in a corresponding $33$ models.
Finally, we test each models' performance on held out samples of completely real, and completely synthetic data to assess their performance and generalization across different training and testing environments.
In each scenario, we analyze if a generator's synthetic data can maintain and/or improve classification accuracy when mixed into the training data at various rates.
Notably, because \system never reproduces any trace identically (Section~\ref{subsec:memorization}) even synthetic flows that are close to real flows in the downstream test set do not constitute train-test leakage, but instead act as supplemental samples for downstream models.

\begin{figure}[t]
  \includegraphics[width=\linewidth]{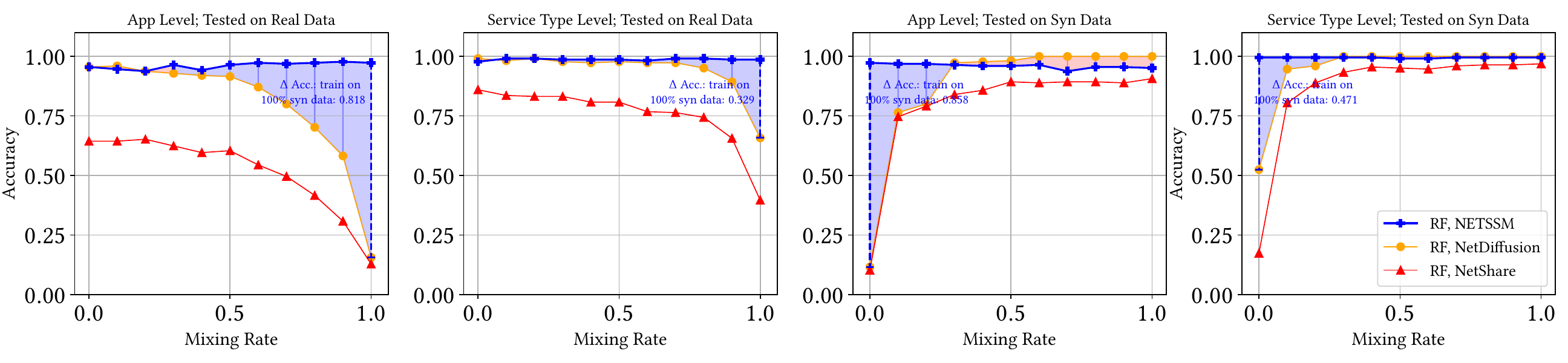}
  \begin{flushleft}
    \vspace{-2mm}
    \makebox[0.155\textwidth]{\hfill \scriptsize (a)}
    \makebox[0.24\textwidth]{\hfill \scriptsize (b)}
    \makebox[0.245\textwidth]{\hfill \scriptsize (c)}
    \makebox[0.245\textwidth]{ \hfill \scriptsize (d)}
  \end{flushleft}

  \caption{\textbf{Performance of random forest classifiers trained on mixed
    real/synthetic data.}
    Models trained on \sysname data perform best across baselines.
  Shading denotes the delta between the next best baseline.}

  \label{fig:ml_simple}
\end{figure}
\subsubsection{Results.}

Figure~\ref{fig:ml_simple} shows the accuracy of RF models trained on the mixed downstream datasets for application and service-type-level traffic classification, and tested on both completely real and synthetic data.
Comprehensive results for other model types, and models trained on non-fine-tuned version \system data are in Appendix~\ref{appendix:downstream}.
We first examine our downstream models performance when tested on completely real data.
Figures~\ref{fig:ml_simple}a and \ref{fig:ml_simple}b visualize the results.
Models trained on \sysname data maintain consistently high classification accuracy in both the absolute and relative cases (as compared to those trained on NetShare and NetDiffusion data), across all mixing rates, even when the training set consists entirely of synthetic data.
We observe substantial accuracy gains of ${\sim}82\%$ and ${\sim}33\%$ for application and service-type-level classification tasks respectively, when synthetic data constitutes 100\% of the training set.
Testing our downstream models on completely synthetic data yields similar results, as shown in Figures~\ref{fig:ml_simple}c and \ref{fig:ml_simple}d.
Models trained on \sysname data consistently achieve high accuracy; at least $0.94$ for application-level classification and $0.99$ for service-type-level classification, regardless of mixing rate.
This represents improvements of ${\sim}86\%$ and ${\sim}47\%$ in either task over the next best synthetic data generator.

\subsection{Semantic Similarity}\label{subsec:semantic_sim}

The existing measures of statistical similarity and downstream utility for network data generators are largely motivated by how well these synthetic data can improve downstream ML-for-networking model performance.
However, raw packet generators introduce new challenges not captured by these metrics.
While a synthetic trace may have both high traffic attribute and header-level similarity, these measures do not reflect the \textit{longitudinal} quality of its contents.
To illustrate, consider a synthetic trace that contains communication between two desired endpoints, but the contained setup and progression between flows is incorrect or out-of-order.
Here, while the five attributes we evaluate in Section~\ref{subsubsec:sfe_content_results} (source/destination IP and port, protocol) would have high statistical similarity, this traffic may not be representative to replace real-world data.

To this end, we evaluate \system's ability to produce \textit{semantically similar} synthetic network traffic. Specifically, we generate both single-flow and \textit{multi-flow} (\ie comprised of more than one connection/five-tuple) Netflix and YouTube video streaming traffic using new \system models further detailed in this section.
We select the workload of video streaming as it contains well-defined patterns which can easily be deemed correctly/incorrectly modeled.
We choose to inspect the multi-flow granularity to examine if the synthesized inter-flow interactions may positively influence the overall fidelity of the trace, and for novelty -- no existing traffic generator can generate multi-flow traffic.
We then evaluate these traces to examine whether they capture implicit characteristics for a given networked communication workload.
To do so, we examine communication between end hosts and synthetic Netflix/YouTube video streaming servers in our generated traces, analyze the attributes of their downloaded video segments, and verify that they reflect the qualities found in real traffic (Section~\ref{subsubsec:semantic_sim_segments}). In our generated multi-flow traces, we further sequentially visualize their segment sending patterns and find that \system's synthetic data captures sending behavior that mimics progression of a real-world video streaming workload.

In all of these analyses, it is not our goal to declaratively state that multi-flow generation is superior to single-flow generation, or vice versa. This would require further work evaluating \system on many additional workloads.
Instead, we want to understand the characteristics of traffic that display higher fidelity (in regard to ground truth traffic) when synthesized in either generation granularity, to better inform how \system can most effectively be used. We provide a recap and further discussion on this point and the results of our analysis in Section~\ref{sec:discussion}.

\subsubsection{Setup.}

We train four \system models, two for Netflix traffic and two for YouTube traffic, on single-flow and multi-flow traffic for Netflix and YouTube video streaming sessions, respectively. Specifically, we use the traces collected by Bronzino~\etal~\cite{bronzino2019inferring} to train our Netflix \system model and the traces from Gutterman~\etal~\cite{gutterman2019requet} to train our YouTube \system model. Table~\ref{tab:dataset_overview} provides an overview of both datasets.
We also note that the video streams contained in our Netflix traces exclusively use TCP-based Dynamic Adaptive Streaming over HTTP~\cite{sodagar2011dash} (DASH), while the video streams in YouTube traces use both TCP and QUIC/UDP-based DASH.

\noindent\textbf{Training.}
For the multi-flow models, we keep all captures in their original, multi-flow state (\ie do not split captures to their comprising single-flows, keep both UDP and TCP traffic), but perform all other pre-processing identically as described in Section~\ref{subsubsec:sfe_setup}.
This results in training datasets of $5{,}882$ and $619$ multi-flow samples for Netflix and YouTube, respectively.
For the single-flow models, we follow the procedure detailed in Section~\ref{subsubsec:sfe_setup} (splitting the original, multi-flow traces into their numerous individual flows), but also filter out training samples comprising fewer than $10{,}000$ packets (a value chosen for reasons described in the following paragraph).
We do this to ensure our single-flow models only learn from the flows which likely correspond to video segment downloads, to provide a fairer comparison against the multi-flow models.
This results in training datasets of $5{,}895$ and $791$ single-flow samples for Netflix and YouTube, respectively.
We pre-train all models using the same training hardware, parameters and input size (samples of $100{,}000$ tokens in length) as described in Section~\ref{subsubsec:sfe_setup} for $30$ epochs.

\noindent\textbf{Generation.}
We then use the models to generate synthetic, single and multi-flow video streaming traffic for both Netflix and YouTube.
We begin the process of creating prompts for generation by filtering the PCAPs in each of the four models' respective training sets ($2 \text{ applications} \times 2 \text{ granularities}$: Netflix single-flow, Netflix multi-flow, YouTube single-flow, YouTube multi-flow) to find the captures with size $\geq 10$MB, likely representative of downloading video streaming content.
We empirically observe that in the multi-flow traces, the video stream segment download patterns described in the previous section for our training data become discernible after $2{,}250$ packets for Netflix, and after $500$ packets for YouTube.
In the single-flow traces, we found these offsets to be $200$ and $25$ packets for Netflix and YouTube respectively.
Accordingly, we use these lengths in our prompts to ``bootstrap'' generation, creating prompts comprised of the tokenized representations of the first corresponding $n \in \{2250, 500, 200, 25\}$ packets from the respective ground truth trace, for each of the four application/granularity pairs.
We create $400$ prompts total from $100$ traces randomly selected from each of the four filtered trace sets.
We then generate one synthetic trace for each prompt, each of length $10{,}000$ packets, for a total of $400$ PCAPs ($100$ for each granularity/application pair). Appendix Section~\ref{sec:gen_params} details the generation hyperparameters for each granularity/application pair.
We choose to generate only $100$ synthetic traces of $10{,}000$ packets each to balance evaluating \system's generation expressiveness over a sufficiently long context, and computational constraints -- each trace takes approximately $20$ minutes to generate, making full-dataset evaluation prohibitive (generating all synthetic traces for our $400$ prompts took approximately five and a half days).
In our below analyzes, we compare the generated traces against their ground truth counterparts, truncated to a matching length of $10{,}000$ packets.

\subsubsection{Results.}
\label{subsubsec:semantic_sim_segments}

We evaluate \system's ability to generate traces that capture the semantics and session dynamics of application-level streaming traffic, for each of the four single/multi-flow and Netflix/YouTube models.
Concretely, for each model's generated traces, we perform one-to-one comparison of a synthetic trace with the original trace whose first $n$ packets were used to prompt its generation, and compare the distributions of their quantities and sizes. Specifically, we infer the DASH \textit{video content segments} found in both the ground truth Netflix/YouTube traces, and the synthetic traces generated by \system.
We use two different definitions of a segment for Netflix and YouTube, respectively, both matching the definition provided in the corresponding original work for either dataset.
For Netflix, we initialize a segment for any uplink packet with non-zero payload, whose destination IP address corresponds to an address received in answers to DNS requests for Netflix domains (\ie \texttt{nflxvideo}, \texttt{netflix}, \texttt{nflxso}) sent at the beginning of a trace.
Subsequent downstream traffic increments the size of the segment.
For YouTube, we initialize a segment for any uplink packet with payload greater than $300$ B, and further only consider it an audio/video segment if it has a final size of at least $80$ KB. Notably, Bronzino~\etal find that Netflix traffic ``downloads, on average, four video segments and one audio segment'' at a given time using many parallel flows, while Gutterman~\etal report that ``for most of [their] dataset, for a given session, audio and video chunks are transmitted from one server.''

We extract these segments from the ground truth real, and synthetic data.
To extract segments from synthetic Netflix traces, we use the IP subnets for Netflix domains found in the ground truth addresses to filter for video stream content, as our generated traces do not contain the DNS payload to perform the same procedure.
All other extraction logic follows as previously described.
We extract YouTube segments from our synthetic traces exactly as previously described.
We then analyze the one-to-one differences in video segment download behavior between synthetic traces, and their corresponding real-world trace used to prompt generation.

\begin{table}[!t]
  \rowcolors{2}{}{lightgray!35}
  \centering

  \caption{\textbf{Statistical and distributional comparisons of video streaming segment downloads.} Comparison between ground truth and corresponding synthetic \system traces.}

  \label{tab:segment_analysis}
  \vspace{-3mm}
  \resizebox{1.\textwidth}{!}{
    \begin{threeparttable}
      \begin{tabular}{p{6em}p{10em}S[table-format=2.2,table-number-alignment=right]S[table-format=2.2,table-number-alignment=right]S[table-format=3.2,table-number-alignment=center]|S[table-format=1.2,table-number-alignment=center]S[table-format=1.2,table-number-alignment=center]|S[table-format=2,table-number-alignment=center]S[table-format=1.2,table-number-alignment=center]|S[table-format=1.2,table-number-alignment=center]|S[table-format=2.1,table-number-alignment=right]}
        \toprule
        \multicolumn{1}{@{}p{6em}}{\multirow{2.25}{*}{\sc{Comp. w/ GT}}} & \multirow{2.25}{*}{\sc{Evaluation}}&
        \multicolumn{3}{c}{\sc{Statistical Measures}}  &
        \multicolumn{2}{c}{\sc{K–S Test}} &
        \multicolumn{2}{c}{\sc{Anderson-Darling Test}}  &
        \multicolumn{1}{c}{\sc{KL Divergence}} & \multicolumn{1}{c}{\sc{EMD}} \\
        \cmidrule[0.5pt](rl){3-5}
        \cmidrule[0.5pt](rl){6-7}
        \cmidrule[0.5pt](rl){8-9}
        \cmidrule[0.5pt](rl){10-10}
        \cmidrule[0.5pt](rl){11-11}
        & & {\sc{Mean $\Delta$}} & {\sc{Median $\Delta$}} & {\sc{Std.
        Dev. $\Delta$}} & {\sc{Stat. $\downarrow$}} & {\sc{p-value
        $\uparrow$}} & {\sc{Stat. $\downarrow$}} & {\sc{p-value
        $\uparrow$}} & {\sc{Stat. (nats) $\downarrow$}} & {\sc{Dist.
        $\downarrow$}} \\
        \midrule

        \rowcolor{white} \multicolumn{11}{@{}l}{\sc{Netflix}}\\
        \midrule

        % NetShare SF & Raw Size & -5086.94  & -7892.77  & 5289.94 & 0.84 & 0.00 & 39.30 & 0.00 & 15.24 & 7246.59 \\
          \system SF & Raw Size & 1.87 & 1.68 & 104.39 & 0.31 & 0.04 & 3.40 & 0.03 & 2.06 & 95.04 \\
          \system MF & Raw Size & -1.09 & 0.00 & 193.91 & 0.21 & 0.03 & 6.86 & 0.01 & 1.14 & 72.94 \\
        \midrule

        % NetShare SF & Avg. Size/Flow & -101.25  & -89.77  & -39.68  & 1.00 & 0.00 & 60.90 & 0.00 & 20.50 & 101.25 \\
          \system SF & Avg. Size/Flow & -0.16 & -0.30 & 38.73 & 0.36 & 0.55 & 0.98 & 0.46 & 3.87 & 57.90 \\
          \system MF & Avg. Size/Flow & -2.27 & -0.66 & 71.73 & 0.22 & 0.82 & 0.29 & 0.73 & 2.67 & 30.75 \\

        \midrule
        % NetShare SF & \# Segments/Flow & 127.28 & -63.50 & 385.99 & 0.60 & 0.00 & 23.80 & 0.00 & 14.13 & 200.75 \\
          \system SF & \# Segments/Flow & 1.69 & 2.00 & 1.93 & 0.48 & 0.19 & 3.00 & 0.06 & 11.37 & 3.95 \\
          \system MF & \# Segments/Flow & -42.24 & 0.00 & 6.22 & 0.17 & 0.95 & 0.41 & 0.76 & 2.97 & 8.04 \\
        \midrule

        \rowcolor{white}\multicolumn{11}{@{}l}{\sc{YouTube}}\\
        \midrule

        % NetShare SF & Raw Size & 14661.37 & 13835.62 & 5714.77 & 0.94 & 0.00 & 14.91 & 0.00 & 15.63 & 14661.37 \\
        \system SF & Raw Size & 10.89 & 431.01 & 41.76 & 0.57 & 0.19 & 3.10 & 0.02 & 12.01 & 592.42 \\
        \system MF & Raw Size & 1.71 & 168.72 & 71.47 & 0.50 & 0.22 & 1.99 & 0.06 & 10.06 & 430.13 \\
        \midrule

        % NetShare SF & Avg. Size/Flow & -14.72  & 50.59  & -295.68  & 0.45 & 0.16 & 0.93 & 0.14 & 7.81 & 229.35 \\
        \system SF & Avg. Size/Flow & -78.43 & 483.07 & {---} & 1.00 & 1.00 & {---} & {---} & 19.56 & 666.35 \\
        \system MF & Avg. Size/Flow & -269.59 & -6.92 & {---} & 0.50 & 0.83 & {---} & {---} & 18.30 & 277.50 \\
        \midrule

        % NetShare SF & \# Segments/Flow & 35.68 & 32.00 & 15.02 & 0.98 & 0.00 & 17.49 & 0.00 & 18.47 & 35.68 \\
        \system SF & \# Segments/Flow & 6.01 & 7.00 & {---} & 1.00 & 0.67 & {---} & {---} & 24.23 & 7.30 \\
        \system MF & \# Segments/Flow & 0.57 & 0.00 & {---} & 0.50 & 1.00 & {---} & {---} & 11.45 & 4.04 \\
        \bottomrule
      \end{tabular}
      \begin{tablenotes}
      \item Values for the statistic, p-value, and distance are the
        median values.\hfill GT := ground truth, SF := single-flow, MF := multi-flow.
      \end{tablenotes}
    \end{threeparttable}
  }
\end{table}

%
%========================================================================================================================
%STATISTICAL COMPARISON METRICS - SYNTHETIC VS GROUND TRUTH
%========================================================================================================================
%
%Aspect                     Mean Δ     Median Δ   Std.Dev. Δ     K-S Stat    K-S p-val     A-D Stat    A-D p-val       KL Div          EMD
%------------------------------------------------------------------------------------------------------------------------------------------------------
%Average Size              -101.25       -89.77       -39.68         1.00         0.00        60.90         0.00        20.50       101.25
%Raw Size                 -5086.94     -7892.77      5289.94         0.84         0.00        39.30         0.00        15.24      7246.59
%Number of Segments         127.28       -63.50       385.99         0.60         0.00        23.80         0.00        14.13       200.75
%======================================================================================================================================================
%
%

\paragraph{Segment Attributes.}
We compare the distributions and summary statistics for segments in regard to 1) raw sizes of all downloaded segments 2) average segment size per flow, and 3) number of segments downloaded per flow, for each ground truth and \system-generated trace pair, for each generation granularity and application. Table~\ref{tab:segment_analysis} provides the summary statistics and results of analysis using various standard statistical measures -- the two-sample Kolmogorov-Smirnov (K-S) and Anderson-Darling tests, Kullback-Leibler (KL) divergence, and the earth mover's distance (EMD) -- for each evaluation. In the \textsc{\small Mean $\Delta$} and \textsc{\small Median $\Delta$} summary statistics, $\Delta \coloneq median_{GT_i}(\texttt{\small eval}) - median_{Gen_i}(\texttt{\small eval})$, where $i \in [1, 100]$, and $GT$ and $Gen$ denote ground truth data and generated data, respectively.
To contrast, \textsc{\small Std.
	Dev.} $\Delta \coloneq median\left(\sigma_{GT_i}(\texttt{\small eval}) - \sigma_{Gen_i}(\texttt{\small eval})\right)$ where $i \in [1, 100]$.
Across statistical measures, smaller values for the statistic or distance and larger $p$-values suggest higher distributional similarity.
We provide additional visualizations that compare other sampled synthetic/ground truth distribution examples in Appendix~\ref{appendix:segment_analysis}.

We observe that as an aggregate across all Netflix traces, \system's synthetic traces of both granularities contain segment downloads whose distribution patterns are similar to the ground truth.
Comparing single and multi-flow traces, we observe that multi-flow traces are more similar to their ground truth counterparts.
While the similar summary statistic values for either granularity are largely comparable (except for \textsc{\small Std.
	Dev.} $\Delta$ and number of segments downloaded per flow), the multi-flow traces have markedly more similar distributions to the ground truth than the single-flow traces.
This is evidenced by the generally lower K-S and Anderson-Darling test statistic and EMD distance values, suggesting large overlap, with additionally larger p-values, in all evaluations except for raw segment size. In the raw segment size evaluation, \system traces of either generation granularity have low median p-values, with the corresponding statistic for the K-S test, KL divergence, and Anderson-Darling test being low, low, and high, respectively.
This suggests that while the general distribution for the traces may be similar, there exist differences in tail values for segment sizes that the traces do not reflect.
This is likely explainable by the nature of the training data.
Because downloading video segments is a largely stable workload across the network conditions in our dataset, our models learn to generate traces that predominantly reflect this norm, with tail segment download behavior less prevalent.

Examining \system-generated YouTube traces, we observe less positively conclusive results.
We omit calculating the standard deviation and two-sample Anderson-Darling test for the average segment size and number of segments per flow, as the ground truth behavior of YouTube traffic is downloading only from one server.
We observe in single-flow generation that for both raw segment size and average segment size per flow, the mean $\Delta$ is close to, or smaller than the ground truth respectively, while the median $\Delta$ is consistently larger. This suggests that in this granularity the majority of downloading flows \system generates typically download segments whose sizes are larger than normally observed in the ground truth, but whose remaining downloaded segments are substantially smaller.
Multi-flow generation displays similar behavior in size of raw segments, but appears to generally synthesize flows whose average segment size is very close to the ground truth.
Unfortunately, it at times appears to ``hallucinate'' numerous outlier flows which download significantly smaller segments, straying from typical YouTube behavior (sequential segment downloads using only a single flow/from one server) and resulting in the substantially negative mean $\Delta$.
The remaining statistical measures are similarly less conclusive, likely for the same reason.
Though we can attempt to ``guide'' \system towards generating traces with only a single dominant flow via generation hyperparameters (\ie influencing token selection), these mechanisms do not ensure that this is adhered to (in either generation granularity). We expect enforcing this constraint on \system would significantly improve its performance for modeling YouTube streaming traffic, and other predominantly single-flow workloads.

\begin{figure}[t]
  \centering
  \begin{subfigure}[b]{.5\textwidth}
    \centering
    \includegraphics[width=\linewidth]{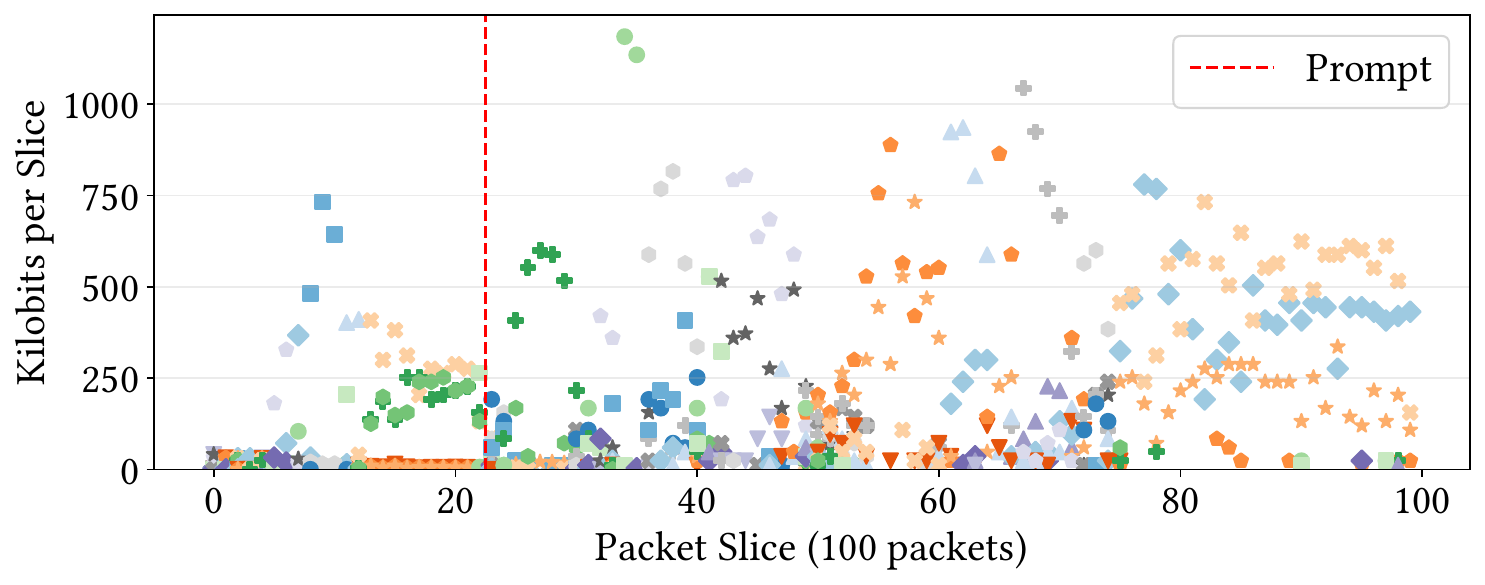}
    \caption{\system-generated Netflix trace.}
    \label{fig:throughput_analysis_a}
  \end{subfigure}\hfill
  \begin{subfigure}[b]{0.5\textwidth}
    \centering
    \includegraphics[width=\linewidth]{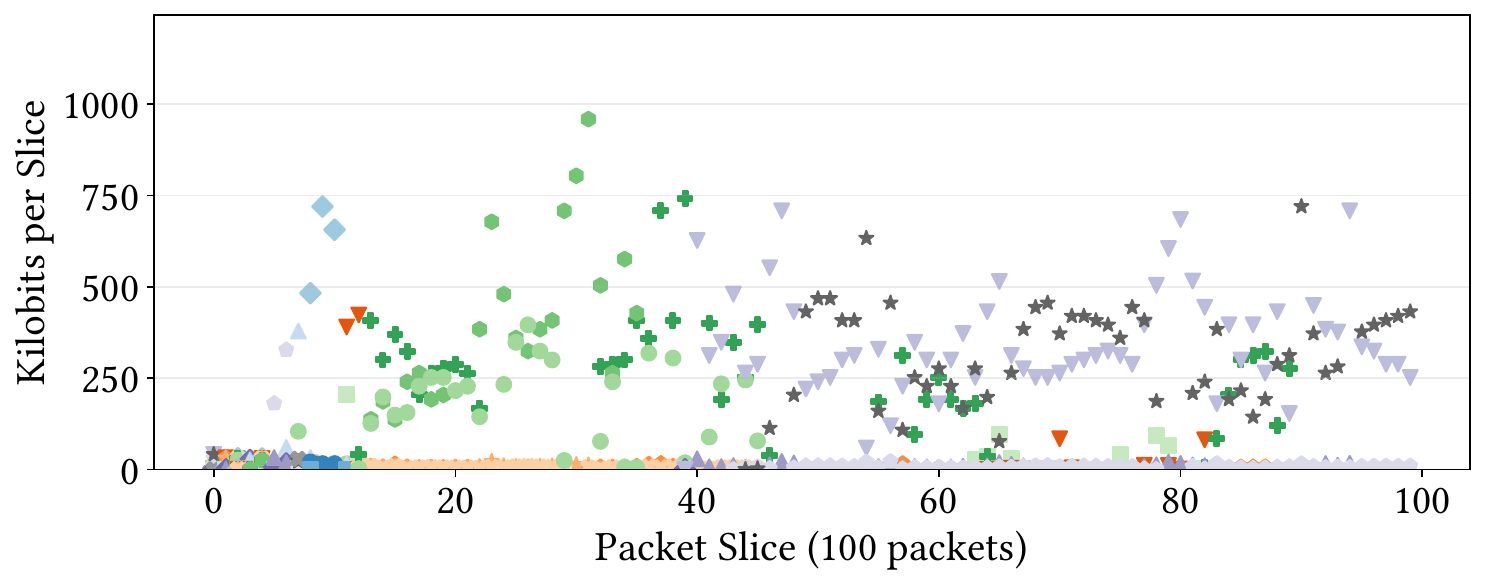}
    \caption{Ground truth Netflix trace.}
    \label{fig:throughput_analysis_b}
  \end{subfigure}\hfill
  \caption{\textbf{Comparison of throughput (synthetic vs. corresponding ground truth trace).}
    Each point's color/shape combination denotes a unique flow.
  Color/shape combinations are not shared between \ref{fig:throughput_analysis_a}/\ref{fig:throughput_analysis_b}.}
  \label{fig:throughput_analysis}
\end{figure}

\paragraph{Sequential Sending Patterns.}\label{para:seqsending}
We next evaluate if the multi-flow traces generated by \system indeed capture the video segment send/download patterns found in real traffic.
This allows us to determine if the events ``behind'' the previous distributional analysis are real-world consistent.
To do so, we plot the throughput of traces based on their comprising flows, as a function of packet order. We do this to present a more direct evaluation of whether \system meaningfully models networked communication over the span of its generation, as timestamps are not truly generated by \system, but assigned post hoc (Section~\ref{subsubsec:generation}).
For a given trace, we partition the trace into slices of $100$ packets.
We then assign the packets in each slice to their corresponding five-tuple flow, and sum the size of all packets in a flow to obtain the throughput in kilobits/slice for that flow.
Figure~\ref{fig:throughput_analysis} visualizes this throughput for both the ground truth (truncated to a matching $10{,}000$ packets) and generated traces for a sample Netflix trace pair.
Appendix Figure~\ref{fig:throughput_extra} contains additional visualizations for either application.
In both \system-generated Netflix and YouTube traces, we see dominant flows that appear empirically similar to the behavior described in the original works (typically three to five active flows for Netflix, one for YouTube).
Unfortunately as previously mentioned, a notable amount of small ``hallucinated'' flows for YouTube traffic are also present, resulting in these traces deviating from the ground truth behavior.

We quantify our analysis by comparing the aggregate throughputs of the generated and ground truth pairs.
We compare aggregate throughput to allow communication between synthetic five-tuples (\ie not present in the ground truth) that reflect the correct sending/receiving behavior of video streaming traffic to count in analysis.
We calculate aggregate throughput by summing the throughput for each flow in a slice, for all slices.
We next calculate two metrics across all synthetic/ground truth pairs: (1) the median Pearson correlation coefficient (PCC) which measures overall alignment of generated and ground truth aggregate throughput, and (2) dynamic time warp (DTW) normalized by the length of the trace which quantifies magnitude-based error while allowing for minor shifts in alignment. We find \system's Netflix traces have moderate positive correlation (PCC=$0.52$), while YouTube traces have weak positive correlation (PCC=$0.31$).
Both results are statistically significant with $p=0.00$.
In magnitude, we find that Netflix and YouTube are $121.45$ and $69.86$ kilobits off from the ground truth at any given moment. Though not optimal, these metrics confirm that while the distribution of segment sizes and downloads synthesized by \system can deviate from the ground truth, the traffic download/sending patterns of the predominant flow(s) are captured.
Thus, it appears that different workloads may likely require a specific generation hyperparameter configuration that best balances generation of realistic segment download distributions alongside the correct sequential communication patterns.

\subsection{Protocol Compliance}\label{subsec:protocol_compliance}
We next evaluate if \system's synthetic traces are ``real-world'' flow and session-compliant, assessing how well they approximate legitimate TCP operation and captures TCP anomalies observed in practice.
Specifically, we compare the TCP state transitions of \system-generated traces for the combined single-flow Multimedia traffic from Sections~\ref{subsec:statistical} and \ref{subsec:ml_task} and single and multi-flow Netflix streaming traffic from Section~\ref{subsec:semantic_sim}, against the behavior of their ground-truth traces.
We also provide comparison against single-flow traces generated by NetDiffusion, without post-generation corrections applied.
TCP is a stateful protocol that requires accurate ordering and flag usage, adherence to handshake procedures, and consistent usage of options.
However, in real network traffic, these behaviors may deviate from RFC specifications for various reasons (\eg middlebox interventions, partial captures).
We parse all traces generated in the previous evaluations using a custom TCP compliance checker that inspects flags, sequence numbers, acknowledgment numbers, and TCP options. Table~\ref{tab:tcp_compliance_split} presents the results of this checker, showing the average percentage change in selected metrics as compared to the ground truth, for both single and multi-flow traces.
We also note several prior and concurrent works that develop model-agnostic methods to ``gate'' synthetic output to be protocol compliant~\cite{hejust,he2025learning}. While relevant, our objective in this analysis, measuring \system's implicit ability to produce TCP-compliant behavior, differs.

\begin{table}[t]
  \centering

  \caption{\textbf{TCP session compliance.} Average percentage change
    in selected metrics as compared to the ground truth, for multi-flow
  \system and single-flow \system and NetDiffusion traces, respectively.}
  % \sysname most reliably captures both protocol specification
  % defined behavior, and real-world anomalies. NetDiffusion, without
  % heuristic fixes, struggles with core protocol states.}

  \vspace{-3mm}
  \label{tab:tcp_compliance_split}
  % \resizebox{0.50\columnwidth}{!}{
  \begin{threeparttable}
    \scriptsize
    \rowcolors{4}{lightgray!35}{}
    \begin{tabular}{l S[table-format=-2.1] S[table-format=-1.1]
      S[table-format=-2.1]}
      \toprule
      \multirow{2.75}{*}{\sc{Metric}}      & \multicolumn{3}{c}{\sc{Model
      (Avg. \%$\Delta$ from ground truth)}}
      \\
      \cmidrule(lr){2-4}
      & \sc{NetSSM (multi-flow)}
      & \sc{NetSSM (single-flow)}    & \sc{NetDiffusion
      (single-flow)}\tnote{\dag}
      \\
      \midrule
      \multicolumn{4}{@{}l}{\itshape TCP session behavior}
      \\
      \midrule
      \texttt{FIN} seen                    & 50.0\%
      & 9.4\%                        & 45.7\%
      \\
      Correct handshakes found             & 0.0\%
      & -5.8\%                       & -70.2\%
      \\
      \texttt{ACK} progress                & -1.0\%
      & -6.5\%                       & -69.6\%
      \\
      \texttt{SEQ} progress                & -1.0\%
      & -8.3\%                       & -68.9\%
      \\
      \texttt{FIN-ACK} observed            & -4.0\%
      & 2.6\%                        & -0.5\%
      \\
      \midrule

      \multicolumn{4}{@{}l}{\itshape Anomalies or deviations in TCP
      behavior}
      \\
      \midrule
      Conflicting flags                    & 57.0\%
      & 0.5\%                        & 34.4\%
      \\
      \texttt{SAck} used w/o \texttt{OK}   & 44.0\%
      & -1.8\%                       & 15.8\%
      \\
      Unexpected \texttt{SYN} after estab. & 6.0\%
      & 0.0\%                        & 0.0\%
      \\
      MSS outside handshake                & 5.0\%
      & 1.4\%                        & -2.0\%
      \\
      WScale outside handshake             & 5.0\%
      & 1.4\%                        & -2.0\%
      \\
      \texttt{RST} in established state    & -16.0\%
      & 0.0\%                        & -35.7\%
      \\
      \bottomrule
    \end{tabular}
    \begin{tablenotes}
      \scriptsize
    \item[] \hfill
      $\dag$ No post-generation fixes applied.
    \end{tablenotes}
  \end{threeparttable}
  % }
\end{table}

We find that \system can produce protocol compliant flows, with the rate of overall compliance being higher for single-flow, as compared to multi-flow \system traces. Single-flow \system traces largely follow the behavior of the ground truth, showing relatively low deltas in expected TCP behavior, and further contain similar rates of anomalies as found in the real data. Multi-flow \system traces are less consistent, showing increased rates of some behaviors (\eg sent \texttt{FIN} packets) and decreased rates of others (\texttt{RST} in established state).
While these are indeed deviations from the training distribution, it is difficult to definitely label this behavior as desirable/undesirable as ground truth PCAPs are often truncated for various reasons unrelated to their comprising communication (\eg capture limits, packet loss, monitoring placement).
Multi-flow traces also display higher rates of some anomalous behavior (\eg conflicting flags).
We analyze the traces corresponding to this behavior and find that \system appears to at times merge consecutive flag states.
Without post-generation correction, single-flow NetDiffusion traces often are not protocol compliant.

\subsection{Memorization Analysis}\label{subsec:memorization}

We verify that \system learns from, rather than memorizes its training data by performing three analyzes on all combined single and multi-flow traces generated in our prior evaluations: (1) one-to-one byte-wise comparison of packets, (2) an approximate matching comparison of packets based on the normalized Hamming distance for each synthetic packet to its nearest neighbor (NN) in the ground truth trace, and (3) a diversity ratio we define as the mean pairwise distance using the same normalized Hamming distance for all synthetic packets, divided by the mean pairwise distance found in the ground truth trace. Appendix Table~\ref{tab:mem_overall} provides detailed results of our analysis.
For (1), we find that on average, only $2.35\%$ of packets are identical per trace. Further, we verify that this percentage corresponds identically to the packets used for prompting \system, accounting for our varying prompt lengths.
In all other differing packets, the average percentage of differing bytes is $22.27\%$, with most differences largely manifesting in fields non-sessional to flow state or protocol compliance.
For (2), we find only 3.83\% of packets lie in a 5\% distance of a ground truth packet, and that this value scales with the distance threshold.
We also run this analysis across different sequential bins of indices (0--10, 10--50, and 50--100 packets) and find that the average NN distance ranges from 0.128 to 0.223, indicating that \system learns deterministic setup phases from its prompt, but generates more varied content as the session progresses.
Finally, we compute a diversity ratio of $0.53$, suggesting that \system produces more closely clustered samples than ground truth traffic.
For applications requiring broader behavioral coverage, generation hyperparameters can be tuned to balance fidelity and diversity.

\section{Discussion, Limitations, and Future Work}\label{sec:discussion}

\noindent\textbf{Improving timestamp generation.}
Currently in \system, timestamps are not learned, but sampled from the distribution of a ground truth capture.
This is not ideal for two reasons.
First, it does not ensure sequential-temporal correlated key events are accurately represented in synthetic traces.
Timestamps assigned to packet may fail to faithfully reflect the true dynamics of the key events which they correspond to.
For instance, if \system was trained on traces containing communication between three endpoints: two on the same local network and one geographically distant, higher latency timestamps of packets to/from the third endpoint could be assigned to communication between the local endpoints, and vice versa.
Second, it requires a \system user retain real data to sample from. This can be especially limiting in exporting trained \system models in environments where the sharing of any data (either raw or derived) is not allowed.
In such cases, \system can still generate PCAP files without sampling timestamps, but the utility of the resulting synthetic traces may substantially decrease, particularly and intuitively when modeling workloads and/or behaviors where time is a large, dependent variable (\eg buffer drain in video streaming).

Ideally, timestamps should be modelled in parallel with, and conditionally based on, flow or packet interactions that arise during generation.
A number of challenges makes this particularly difficult.
First, the modalities of packet contents (discrete values for bytes) and timestamps (continuous values for duration since epoch) are not the same, and thus cannot be simultaneously modeled using the same objective function.
During experimentation, we confirmed this challenge when we attempted to modify \system to contain an additional regression modelling component that took as input the most recent generated two consecutive packets, and output the packet IAT.
Despite training this component under various custom loss functions, we were unable to obtain a model that accurately captured IATs.
Instead, the generated IATs roughly regressed to the mean of the trace, despite loss functions placing emphasis on spikes, or other long-tail events.
Second, there exist influences external to the data and communication contained in packets (\eg physical distance, link outages) that may have far greater impact on the timestamp value.
While prior work in the ML-community~\cite{zhang2023mixed,somepalli2021saint} has demonstrated simultaneous generation of both discrete and continuous-typed tabular data, the causal dependencies for these mixed types are wholly contained in each independent sample.
This contrasts with the scope of our modelling, where timestamps have causal dependencies on the external influences mentioned above, not captured in the PCAP data \system learns from.
As such, it seems necessary to in tandem, consider a \textit{third} modality that captures a network's topological characteristics to inform timestamp generation.

An alternative approach may involve a special token which demarcates packet content from discretized (\ie tokenized) representations of the timestamp, allowing for training under the same cross-entropy loss function (though the benefits/detriments of discretizing time must be considered).
Future improvements to \system's timestamp generation and broader efforts to model both packet contents and time in parallel should thus consider how to reason about the different modalities involved, and attempt to incorporate outside influences not present in the capture itself.

\noindent\textbf{Choosing generation granularity (single-flow versus multi-flow).}
\system is the first network traffic generator that can generate raw packet traces comprised of either a single flow, or multiple flows.
However, it is important to consider which granularity is most appropriate when modelling a given networked workload.
An example of this is in Section~\ref{para:seqsending}, where instructing \system to generate a multi-flow trace for YouTube traffic, a workload whose ground truth is predominantly comprised of only a single audio and video downloading flow, results in a substantial number of ``hallucinated flows.'' Alternatively, synthetic traces for Netflix traffic, a workload whose ground truth is predominantly comprised of five flows (one downloading audio and four downloading video) show moderate positive correlation to the ground truth.
We provide this case study to provide more thorough analysis of \system's behavior for generating video streaming traffic, and to view how \system behaves given different sending and receiving dynamics (\ie YouTube's single server audio/video chunk transmission versus Netflix's multiple server transmission).
Intuitively, using a single-flow-trained \system model to learn and generate a predominantly single-flow workload will very likely yield substantially better results.
Taken a step further, it may be more effective to learn from and independently generate multiple key single-flow traces for a given workload, before combining them in a unified, interleaved trace.
However, determining how to order the arrival and interleaving of flows may be a non-trivial task.

\noindent\textbf{Generating and evaluating more diverse network data.}
In this work, we generate single and multi-flow traces for various multimedia traffic. We choose these workloads as they are straightforward, and provide a solid starting point for evaluating \system's synthetic data.
Follow-up work should explore generating more diverse, and/or complicated traffic. One immediate direction is extending \system's pre-processor to parse traces comprised of additional transport and application layer protocols (\eg QUIC, RTP).
This is easily implementable, only requiring writing an additional handler function within our Go parser; all ML-related operation of \system remains the same. Additionally, we find that \system's performance may vary based on generation hyperparameters to best suit a target workload.
As such, additional modelling of different network workloads could help to better understand if different patterns of parameters possibly exist for different traffic.

\section{Conclusion}\label{sec:conclusion}

In this paper, we presented \system, an SSM-based raw packet generator.
\system's sequential, stateful architecture enables it to learn from, and
produce sessions $8\times$ and $78\times$ longer, respectively, than the current
state-of-the-art transformer-based raw packet generator.
This allows it to capture key flow-state-dependent session events at both the single and multi-flow session granularities that only manifest after substantial setup. \system outperforms all previous generators in measures of statistical similarity and as measured by the performance of downstream ML-for-networking models trained on \system data.
We additionally pose a new evaluation of semantic similarity that attempts to better reason about the empirical, practical similarities between synthetic output and real-world network data. We find that \system can capture complex application dynamics of multi-flow networked communication. Finally, we verify that \system's generated traces largely reproduce the TCP-adherent, and anomalous behaviors found in real traffic data.
This paper does not raise any ethical concerns.

\begin{acks}
	We thank our shepherd Alessandro Finamore and our anonymous reviewers for their feedback and suggestions.
This work has been supported by grants from the Agence Nationale de la Recherche (project no. ANR-21-CE94-0001 [MINT]), the National Science Foundation (grant nos. CNS-2334996 and CNS-2319603), and the France and Chicago Collaborating in The Sciences program.

\end{acks}

\bibliographystyle{ACM-Reference-Format}
\bibliography{ref}

@inproceedings{holland2021new,
    title = {New directions in automated traffic analysis},
    author = {Holland, Jordan and Schmitt, Paul and Feamster, Nick and Mittal, Prateek},
    booktitle = {Proceedings of the 2021 ACM SIGSAC conference on computer and communications security},
    pages = {3366--3383},
    year = {2021},
}

@techreport{ietf-opsawg-pcap-06,
    number = {draft-ietf-opsawg-pcap-06},
    type = {Internet-Draft},
    institution = {Internet Engineering Task Force},
    publisher = {Internet Engineering Task Force},
    note = {Work in Progress},
    url = {https://datatracker.ietf.org/doc/draft-ietf-opsawg-pcap/06/},
    author = {Guy Harris and Michael Richardson},
    title = {{PCAP Capture File Format}},
    pagetotal = 10,
    year = 2025,
    month = sep,
    day = 3,
}

@article{zhang2023mixed,
    title = {Mixed-type tabular data synthesis with score-based diffusion in latent space},
    author = {Zhang, Hengrui and Zhang, Jiani and Srinivasan, Balasubramaniam and Shen, Zhengyuan and Qin, Xiao and Faloutsos, Christos and Rangwala, Huzefa and Karypis, George},
    journal = {arXiv preprint arXiv:2310.09656},
    year = {2023},
}

@article{somepalli2021saint,
    title = {Saint: Improved neural networks for tabular data via row attention and contrastive pre-training},
    author = {Somepalli, Gowthami and Goldblum, Micah and Schwarzschild, Avi and Bruss, C Bayan and Goldstein, Tom},
    journal = {arXiv preprint arXiv:2106.01342},
    year = {2021},
}

@inproceedings{gutterman2019requet,
    title = {Requet: Real-time QoE detection for encrypted YouTube traffic},
    author = {Gutterman, Craig and Guo, Katherine and Arora, Sarthak and Wang, Xiaoyang and Wu, Les and Katz-Bassett, Ethan and Zussman, Gil},
    booktitle = {Proceedings of the 10th ACM Multimedia Systems Conference},
    pages = {48--59},
    year = {2019},
}

@article{hejust,
    title = {Just-in-Time Logic Enforcement},
    author = {H{\`e}, Hongyu and Apostolaki, Maria},
}

@article{he2025learning,
    title = {Learning Constraints Directly from Network Data},
    author = {H{\`e}, Hongyu and Jin, Minhao and Apostolaki, Maria},
    journal = {arXiv preprint arXiv:2506.23964},
    year = {2025},
}

@software{Moi_HuggingFace_s_Tokenizers_2023,
    author = {Moi, Anthony and Patry, Nicolas},
    license = {Apache-2.0},
    month = apr,
    title = {{HuggingFace's Tokenizers}},
    url = {https://github.com/huggingface/tokenizers},
    version = {0.13.4},
    year = {2023},
}

@inproceedings{dao2024mamba2,
    title = {Transformers are {SSM}s: Generalized Models and Efficient Algorithms Through Structured State Space Duality},
    author = {Dao, Tri and Gu, Albert},
    booktitle = {Proceedings of the 41st International Conference on Machine Learning},
    pages = {10041--10071},
    year = {2024},
    editor = {Salakhutdinov, Ruslan and Kolter, Zico and Heller, Katherine and Weller, Adrian and Oliver, Nuria and Scarlett, Jonathan and Berkenkamp, Felix},
    volume = {235},
    series = {Proceedings of Machine Learning Research},
    month = {21--27 Jul},
    publisher = {PMLR},
    url = {https://proceedings.mlr.press/v235/dao24a.html},
}

@article{gu2021s4,
    title = {Efficiently modeling long sequences with structured state spaces},
    author = {Gu, Albert and Goel, Karan and R{\'e}, Christopher},
    journal = {arXiv preprint arXiv:2111.00396},
    year = {2021},
}

@article{gu2020hippo,
    title = {Hippo: Recurrent memory with optimal polynomial projections},
    author = {Gu, Albert and Dao, Tri and Ermon, Stefano and Rudra, Atri and R{ \'e}, Christopher},
    journal = {Advances in neural information processing systems},
    volume = {33},
    pages = {1474--1487},
    year = {2020},
}

@article{voelker2019legendre,
    title = {Legendre memory units: Continuous-time representation in recurrent neural networks},
    author = {Voelker, Aaron and Kaji{\'c}, Ivana and Eliasmith, Chris},
    journal = {Advances in neural information processing systems},
    volume = {32},
    year = {2019},
}

@article{gu2023mamba,
    title = {Mamba: Linear-time sequence modeling with selective state spaces},
    author = {Gu, Albert and Dao, Tri},
    journal = {arXiv preprint arXiv:2312.00752},
    year = {2023},
}

@article{kalman1960new,
    title = {A new approach to linear filtering and prediction problems},
    author = {Kalman, Rudolph Emil},
    year = {1960},
}

@inproceedings{chu2024feasibility,
    title = {Feasibility of state space models for network traffic generation},
    author = {Chu, Andrew and Jiang, Xi and Liu, Shinan and Bhagoji, Arjun and Bronzino, Francesco and Schmitt, Paul and Feamster, Nick},
    booktitle = {Proceedings of the 2024 SIGCOMM Workshop on Networks for AI Computing},
    pages = {9--17},
    year = {2024},
}

@inproceedings{Lin_2022,
    series = {WWW ’22},
    title = {ET-BERT: A Contextualized Datagram Representation with Pre-training Transformers for Encrypted Traffic Classification},
    url = {http://dx.doi.org/10.1145/3485447.3512217},
    DOI = {10.1145/3485447.3512217},
    booktitle = {Proceedings of the ACM Web Conference 2022},
    publisher = {ACM},
    author = {Lin, Xinjie and Xiong, Gang and Gou, Gaopeng and Li, Zhen and Shi, Junzheng and Yu, Jing},
    year = {2022},
    month = apr,
    collection = {WWW ’22},
}

@article{qu2024trafficgpt,
    title = {TrafficGPT: Breaking the Token Barrier for Efficient Long Traffic Analysis and Generation},
    author = {Qu, Jian and Ma, Xiaobo and Li, Jianfeng},
    journal = {arXiv preprint arXiv:2403.05822},
    year = {2024},
}

@article{meng2023netgpt,
    title = {Netgpt: Generative pretrained transformer for network traffic},
    author = {Meng, Xuying and Lin, Chungang and Wang, Yequan and Zhang, Yujun},
    journal = {arXiv preprint arXiv:2304.09513},
    year = {2023},
}

@inproceedings{abt2014we,
    title = {Are we missing labels? A study of the availability of ground-truth in network security research},
    author = {Abt, Sebastian and Baier, Harald},
    booktitle = {2014 third international workshop on building analysis datasets and gathering experience returns for security (badgers)},
    pages = {40--55},
    year = {2014},
    organization = {IEEE},
}

@inproceedings{yin2022practical,
    title = {Practical gan-based synthetic ip header trace generation using netshare},
    author = {Yin, Yucheng and Lin, Zinan and Jin, Minhao and Fanti, Giulia and Sekar, Vyas},
    booktitle = {Proceedings of the ACM SIGCOMM 2022 Conference},
    pages = {458--472},
    year = {2022},
}

@inproceedings{lin2020using,
    title = {Using gans for sharing networked time series data: Challenges, initial promise, and open questions},
    author = {Lin, Zinan and Jain, Alankar and Wang, Chen and Fanti, Giulia and Sekar, Vyas},
    booktitle = {Proceedings of the ACM Internet Measurement Conference},
    pages = {464--483},
    year = {2020},
}

@misc{pcapsplitter,
    author = {shramos},
    title = {shramos/pcap-splitter},
    year = {2019},
    howpublished = {\url{https://github.com/shramos/pcap-splitter}},
}

@article{jiang2024netdiffusion,
    title = {NetDiffusion: Network Data Augmentation Through Protocol-Constrained Traffic Generation},
    author = {Jiang, Xi and Liu, Shinan and Gember-Jacobson, Aaron and Bhagoji, Arjun Nitin and Schmitt, Paul and Bronzino, Francesco and Feamster, Nick},
    journal = {Proceedings of the ACM on Measurement and Analysis of Computing Systems},
    volume = {8},
    number = {1},
    pages = {1--32},
    year = {2024},
    publisher = {ACM New York, NY, USA},
}

@article{jiang2025jiti,
    title = {JITI: Dynamic Model Serving for Just-in-Time Traffic Inference},
    author = {Jiang, Xi and Liu, Shinan and Naama, Saloua and Bronzino, Francesco and Schmitt, Paul and Feamster, Nick},
    journal = {Proceedings of the ACM on Networking},
    volume = {3},
    number = {CoNEXT4},
    pages = {1--24},
    year = {2025},
    publisher = {ACM New York, NY, USA},
}

@inproceedings{li2022foap,
    title = {$\{$FOAP$\}$:$\{$Fine-Grained$\}$$\{$Open-World$\}$ android app fingerprinting},
    author = {Li, Jianfeng and Zhou, Hao and Wu, Shuohan and Luo, Xiapu and Wang , Ting and Zhan, Xian and Ma, Xiaobo},
    booktitle = {31st USENIX Security Symposium (USENIX Security 22)},
    pages = {1579--1596},
    year = {2022},
}

@inproceedings{sharma2023estimating,
    author = {Sharma, Taveesh and Mangla, Tarun and Gupta, Arpit and Jiang, Junchen and Feamster, Nick},
    title = {Estimating WebRTC Video QoE Metrics Without Using Application Headers},
    year = {2023},
    isbn = {9798400703829},
    publisher = {Association for Computing Machinery},
    address = {New York, NY, USA},
    url = {https://doi.org/10.1145/3618257.3624828},
    doi = {10.1145/3618257.3624828},
    booktitle = {Proceedings of the 2023 ACM on Internet Measurement Conference},
    pages = {485–500},
    numpages = {16},
    location = {Montreal QC, Canada},
    series = {IMC '23},
}

@article{swann2021tools,
    title = {Tools for Network Traffic Generation--A Quantitative Comparison},
    author = {Swann, Matthew and Rose, Joseph and Bendiab, Gueltoum and Shiaeles , Stavros and Savage, Nick},
    journal = {arXiv preprint arXiv:2109.02760},
    year = {2021},
}

@article{campanile2020computer,
    title = {Computer network simulation with ns-3: A systematic literature review},
    author = {Campanile, Lelio and Gribaudo, Marco and Iacono, Mauro and Marulli , Fiammetta and Mastroianni, Michele},
    journal = {Electronics},
    volume = {9},
    number = {2},
    pages = {272},
    year = {2020},
    publisher = {MDPI},
}

@misc{ciscotrex2023,
    title = "The {CISCO} {TRex} {Tool}",
    year = "2024",
    HowPublished = "\url{https://trex-tgn.cisco.com/}",
    Key = {ciscotrex2023},
    note = "[Online; accessed 31-May-2024]",
}

@inproceedings{lacage2006yet,
    title = {Yet another network simulator},
    author = {Lacage, Mathieu and Henderson, Thomas R},
    booktitle = {Proceedings of the 2006 Workshop on ns-3},
    pages = {12--es},
    year = {2006},
}

@article{henderson2008network,
    title = {Network simulations with the ns-3 simulator},
    author = {Henderson, Thomas R and Lacage, Mathieu and Riley, George F and Dowell, Craig and Kopena, Joseph},
    journal = {SIGCOMM demonstration},
    volume = {14},
    number = {14},
    pages = {527},
    year = {2008},
}

@inproceedings{buhler2022generating,
    title = {Generating representative, live network traffic out of millions of code repositories},
    author = {B{\"u}hler, Tobias and Schmid, Roland and Lutz, Sandro and Vanbever, Laurent},
    booktitle = {Proceedings of the 21st ACM Workshop on Hot Topics in Networks},
    pages = {1--7},
    year = {2022},
}

@article{botta2012tool,
    title = {A tool for the generation of realistic network workload for emerging networking scenarios},
    author = {Botta, Alessio and Dainotti, Alberto and Pescap{\'e}, Antonio},
    journal = {Computer Networks},
    volume = {56},
    number = {15},
    pages = {3531--3547},
    year = {2012},
    publisher = {Elsevier},
}

@inproceedings{nprint,
    author = {Holland, Jordan and Schmitt, Paul and Feamster, Nick and Mittal, Prateek},
    title = {New Directions in Automated Traffic Analysis},
    year = {2021},
    isbn = {9781450384544},
    publisher = {Association for Computing Machinery},
    address = {New York, NY, USA},
    url = {https://doi.org/10.1145/3460120.3484758},
    doi = {10.1145/3460120.3484758},
    pages = {3366–3383},
    numpages = {18},
    location = {Virtual Event, Republic of Korea},
    series = {CCS '21},
}

@misc{rfc7858,
    series = {Request for Comments},
    number = 7858,
    howpublished = {RFC 7858},
    publisher = {RFC Editor},
    doi = {10.17487/RFC7858},
    url = {https://www.rfc-editor.org/info/rfc7858},
    author = {Zi Hu and Liang Zhu and John Heidemann and Allison Mankin and Duane Wessels and Paul E. Hoffman},
    title = {{Specification for DNS over Transport Layer Security (TLS)}},
    pagetotal = 19,
    year = 2016,
    month = may,
}

@misc{rfc8484,
    series = {Request for Comments},
    number = 8484,
    howpublished = {RFC 8484},
    publisher = {RFC Editor},
    doi = {10.17487/RFC8484},
    url = {https://www.rfc-editor.org/info/rfc8484},
    author = {Paul E. Hoffman and Patrick McManus},
    title = {{DNS Queries over HTTPS (DoH)}},
    pagetotal = 21,
    year = 2018,
    month = oct,
}

@article{bronzino2019inferring,
    title = {Inferring streaming video quality from encrypted traffic: Practical models and deployment experience},
    author = {Bronzino, Francesco and Schmitt, Paul and Ayoubi, Sara and Martins , Guilherme and Teixeira, Renata and Feamster, Nick},
    journal = {Proceedings of the ACM on Measurement and Analysis of Computing Systems},
    volume = {3},
    number = {3},
    pages = {1--25},
    year = {2019},
    publisher = {ACM New York, NY, USA},
}

@inproceedings{macmillan2021measuring,
    title = {Measuring the performance and network utilization of popular video conferencing applications},
    author = {MacMillan, Kyle and Mangla, Tarun and Saxon, James and Feamster, Nick},
    booktitle = {Proceedings of the 21st ACM Internet Measurement Conference},
    pages = {229--244},
    year = {2021},
}

@article{sodagar2011dash,
    author = {Sodagar, Iraj},
    journal = {IEEE MultiMedia},
    title = {The MPEG-DASH Standard for Multimedia Streaming Over the Internet},
    year = {2011},
    volume = {18},
    number = {4},
    pages = {62-67},
    doi = {10.1109/MMUL.2011.71},
}

@misc{letsencrypthttps,
    title = {Let's Encrypt Stats},
    author = {Let's Encrypt},
    year = {2024},
    url = {https://letsencrypt.org/stats/},
    note = {Accessed: 2024},
}

@article{singhal2013state,
    title = {State of the Art Review of Network Traffic Classification based on Machine Learning Approach},
    author = {Singhal, Pallavi and Mathur, Rajeev and Vyas, Himani},
    journal = {International Journal of Computer Applications},
    volume = {975},
    pages = {8887},
    year = {2013},
}

@inproceedings{5504793,
    author = {Sommer, Robin and Paxson, Vern},
    booktitle = {2010 IEEE Symposium on Security and Privacy},
    title = {Outside the Closed World: On Using Machine Learning for Network Intrusion Detection},
    year = {2010},
    volume = {},
    number = {},
    pages = {305-316},
    doi = {10.1109/SP.2010.25},
}

@article{paxson1999bro,
    title = {Bro: a system for detecting network intruders in real-time},
    author = {Paxson, Vern},
    journal = {Computer networks},
    volume = {31},
    number = {23-24},
    pages = {2435--2463},
    year = {1999},
    publisher = {Elsevier},
}

@article{baker2004cisco,
    title = {Cisco architecture for lawful intercept in IP networks},
    author = {Baker, Fred and Foster, Bill and Sharp, Chip},
    journal = {Internet Engineering Task Force, RFC},
    volume = {3924},
    year = {2004},
    publisher = {Citeseer},
}

@article{de2023survey,
    title = {A Survey of Public IoT Datasets for Network Security Research},
    author = {De Keersmaeker, Fran{\c{c}}ois and Cao, Yinan and Ndonda, Gorby Kabasele and Sadre, Ramin},
    journal = {IEEE Communications Surveys \& Tutorials},
    year = {2023},
    publisher = {IEEE},
}

@article{2020SciPy-NMeth,
    author = {Virtanen, Pauli and Gommers, Ralf and Oliphant, Travis E. and Haberland, Matt and Reddy, Tyler and Cournapeau, David and Burovski , Evgeni and Peterson, Pearu and Weckesser, Warren and Bright, Jonathan and {van der Walt}, St{\'e}fan J. and Brett, Matthew and Wilson, Joshua and Millman, K. Jarrod and Mayorov, Nikolay and Nelson, Andrew R. J. and Jones, Eric and Kern, Robert and Larson, Eric and Carey, C J and Polat, {\.I}lhan and Feng, Yu and Moore, Eric W. and {VanderPlas}, Jake and Laxalde, Denis and Perktold, Josef and Cimrman, Robert and Henriksen, Ian and Quintero, E. A. and Harris, Charles R. and Archibald, Anne M. and Ribeiro, Ant{\^o} nio H. and Pedregosa, Fabian and {van Mulbregt}, Paul and {SciPy 1.0 Contributors}},
    title = {{{SciPy} 1.0: Fundamental Algorithms for Scientific Computing in Python}},
    journal = {Nature Methods},
    year = {2020},
    volume = {17},
    pages = {261--272},
    adsurl = {https://rdcu.be/b08Wh},
    doi = {10.1038/s41592-019-0686-2},
}

@book{scott2015multivariate,
    title = {Multivariate density estimation: theory, practice, and visualization},
    author = {Scott, David W},
    year = {2015},
    publisher = {John Wiley \& Sons},
}

@article{zhang2024netdiff,
    title = {NetDiff: A Service-Guided Hierarchical Diffusion Model for Network Flow Trace Generation},
    author = {Zhang, Shiyuan and Li, Tong and Jin, Depeng and Li, Yong},
    journal = {Proceedings of the ACM on Networking},
    volume = {2},
    number = {CoNEXT3},
    pages = {1--21},
    year = {2024},
    publisher = {ACM New York, NY, USA},
}
%% If your work has an appendix, this is the place to put it.
\newpage
\appendix
\setcounter{table}{0}
\setcounter{figure}{0}

\section{Comprehensive Results on Downstream
  Utilization}\label{appendix:downstream}
\begin{figure}[h]
  \centering
  \includegraphics[width=\linewidth]{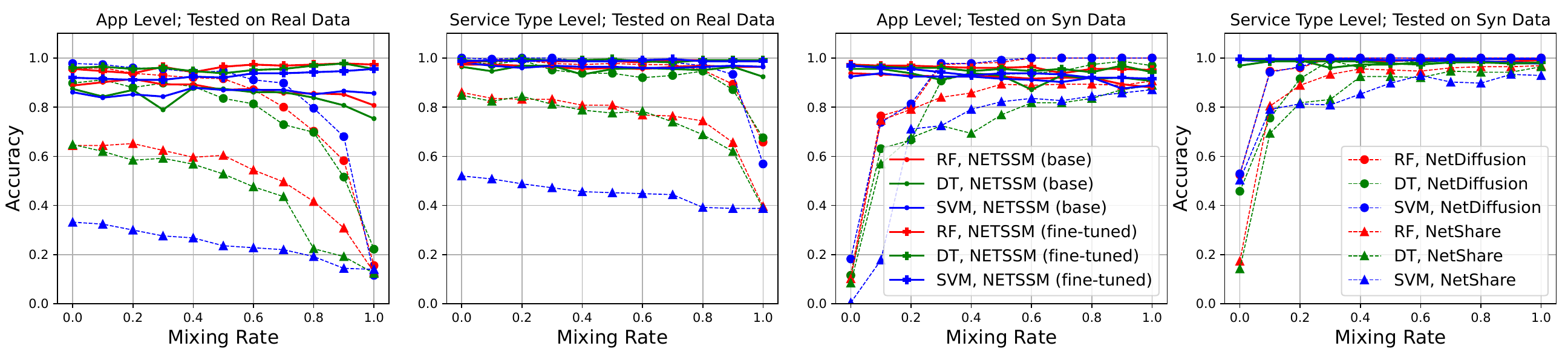}
  \vspace{-7mm}
  \caption{Comparative ML performance across different model choices
  with mixed training data proportions.}
  \label{fig:ml_comprehensive}
\end{figure}

% \begin{figure}[h]
  %   \centering
  %   \includegraphics[width=\linewidth]{evaluation_compare_comprehensive2_3.pdf}
  %   \vspace{-3mm}
  %   \caption{Comparative ML performance across different model choices
  %     with mixed training data proportions (Skipping first packet for
  %   \sysname-generated traces).}
  %   \label{fig:ml_comprehensive_no_seed}
  % \end{figure}

% \section{Aggregate TCP Options Analysis of Synthetic Data}
%
% \input{tables/tcp_options.tex}
%
% Table~\ref{tab:tcp_options} shows that \system's multi-flow traces exhibit
% significantly higher raw counts for MSS, WScale, and \texttt{SAckOK} compared
% to their real counterparts, potentially reflecting how multi-flow generation
% amplifies the number of handshake or handshake-like packets emitted by \system.
% By contrast, Timestamp usage is extremely frequent for both real and synthetic
% traces, although single-flow real PCAPs exceed synthetic usage by a large
% margin. These differences underline that captured durations or concurrency
% levels can strongly influence aggregated option counts, making perfect numeric
% alignment a challenging goal. Rather than detracting from fidelity, these
% variants illustrate that \system can produce either condensed or expanded views
% of typical network conditions, including advanced TCP features that manifest
% over extended sessions.

\section{Additional Video Streaming Segment
  Results}\label{appendix:segment_analysis}
% 5f43a2eb8f7fd1628505f72f8c54f536
% 7ecebbc9a5956999207f375096bb73ee

% 58087ec9a7c68bfa1a98b1bb4652fcc0
% 5c14027a40c1189a46ad4962f1dfda13
% 938d2186acb2f611ab3e10834d52df3e
% 30cc5499f5058f639ebb272ebf3fb56a

The scenarios shown in all figures below have the following ground truth data bit rates: (1) $554$ kbps, (2) $1{,}366$ kbps, (3) $2{,}726$ kbps, (4) $2{,}460$ kbps, (5) $1{,}361$ kbps, and (6) $1{,}450$ kbps.

\begin{figure}[h]
  \centering
  \begin{subfigure}[b]{0.208\textwidth}
    \centering
    \includegraphics[width=\linewidth]{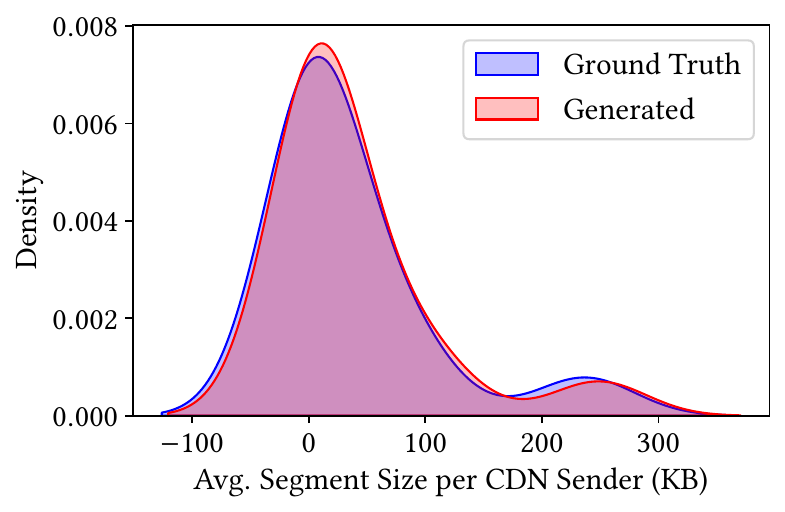}
    \caption{}
    \label{fig:segment_analysis_a}
  \end{subfigure}\hfill
  \begin{subfigure}[b]{0.208\textwidth}
    \centering
    \includegraphics[width=\linewidth]{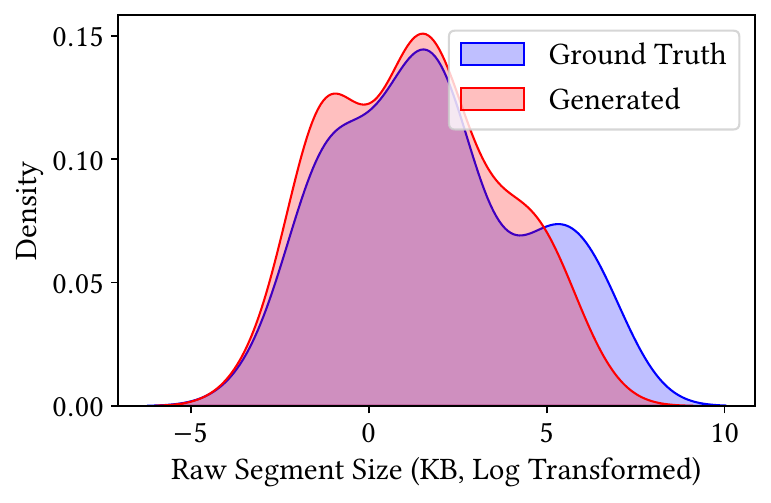}
    \caption{}
    \label{fig:segment_analysis_b}
  \end{subfigure}\hfill
  \begin{subfigure}[b]{0.205\textwidth}
    \centering
    \includegraphics[width=\linewidth]{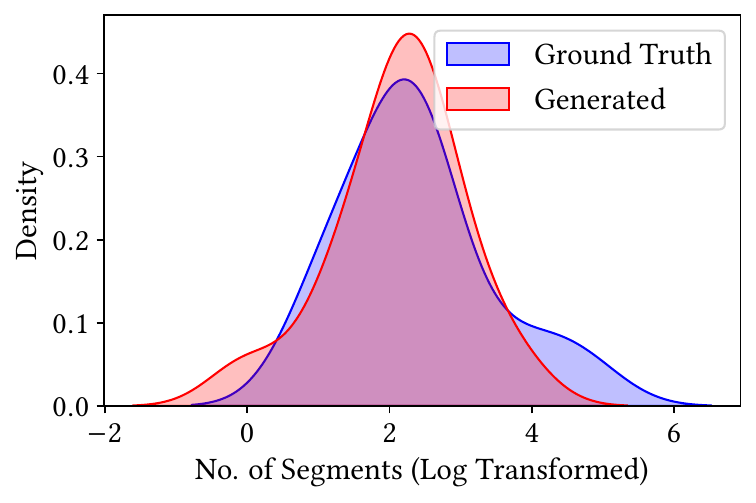}
    \caption{}
    \label{fig:segment_analysis_c}
  \end{subfigure}\hfill
  \begin{subfigure}[b]{0.177\textwidth}
    \centering
    \includegraphics[width=\linewidth]{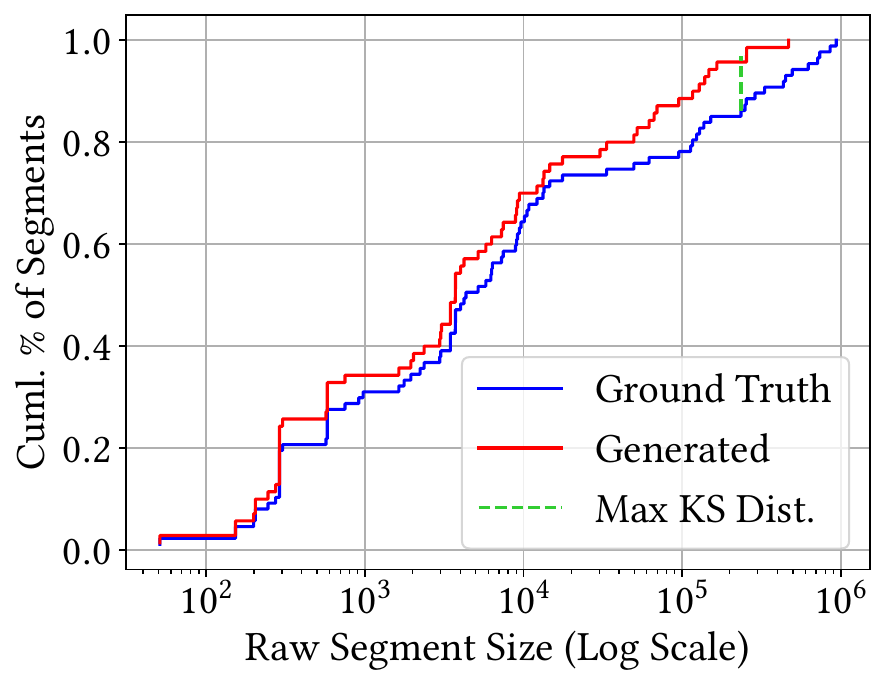}
    \caption{}
    \label{fig:segment_analysis_d}
  \end{subfigure}\hfill
  \begin{subfigure}[b]{0.177\textwidth}
    \centering
    \includegraphics[width=\linewidth]{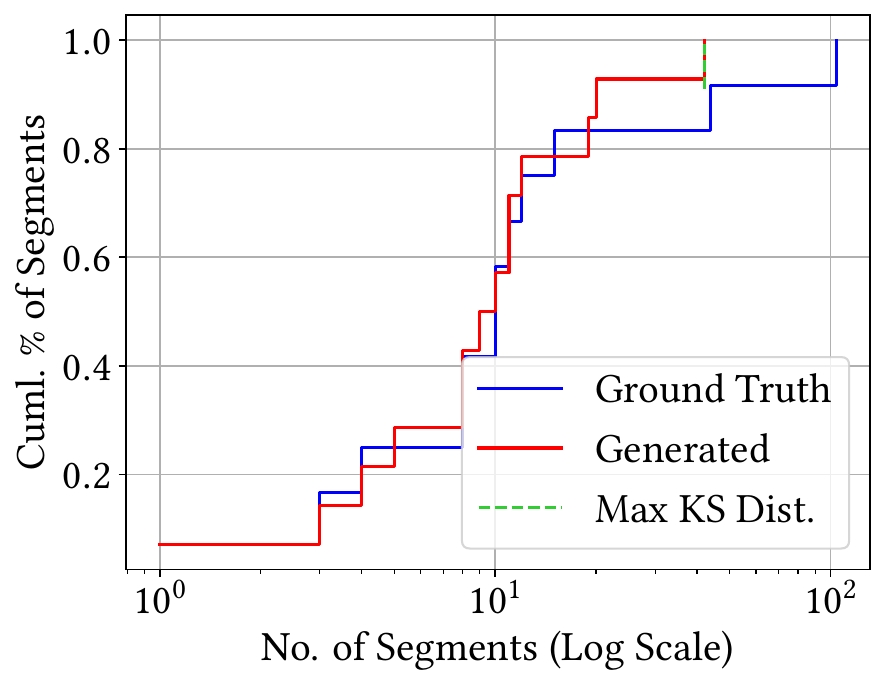}
    \caption{}
    \label{fig:segment_analysis_e}
  \end{subfigure}
  \caption{\textbf{Distributions for downloaded segments.}
    KDE (log-transformed) and ECDF (non-log-transformed, displayed on log scale) plots for the number and size of downloaded segments sent per sender.
    The
    ground truth trace has a data bit rate of $1{,}366$ kbps. \system's
  distributions overlap significantly with the real data.}
  \label{fig:segment_analysis}
\end{figure}

\noindent\textbf{Downloaded Segment Sizes.}
Figure~\ref{fig:segment_analysis} shows applying kernel density estimation (KDE) to the average segment sizes per sender (\ref{fig:segment_analysis_a}) and log-transformed sizes of all raw segment sizes (\ref{fig:segment_analysis_b}), and the empirical cumulative distribution function (ECDF) for raw segment sizes (\ref{fig:segment_analysis_d}) for a ground truth and corresponding generated trace.
% We choose to apply log
% transformation to the raw segment sizes to better visualize the two
% distributions, as both the ground truth and generated distributions of sizes are
% positively right-skewed with a higher volume of smaller segments (\ie
% corresponding to session setup) than large segments (\ie corresponding to actual
% video content download) present. We use log scale in the ECDF plots for the same
% reason.
All KDE plots are created using a Gaussian kernel with (ground truth, generated) bandwidths of$(40.62, 39.08)$, and $(1.07, 0.99)$ for Figures~\ref{fig:segment_analysis_a}, and \ref{fig:segment_analysis_b} respectively, chosen using Scott's rule of thumb in the Python \texttt{scipy} library~\cite{scott2015multivariate,2020SciPy-NMeth}.
These figures well illustrate the similarity in downloaded segment sizes, where Observing Figures \ref{fig:segment_analysis_a} and \ref{fig:segment_analysis_b}, clear overlap exists between the segment sizes of the ground truth and synthetic data, even when considering instances of larger tail values.
Similarly, in the Figure~\ref{fig:segment_analysis_d} ECDF, the generated trace segment sizes overlap with the ground truth, as illustrated by similar magnitudes in the 25th, 50th, and 75th quartiles: $(580.00, 4344.00, 41564.00)$ KB for ground truth and $(369.50, 3721.00, 14349.50)$ KB for generated, respectively.
As such, we see that \system generates traces with similar size magnitudes across all segments, and with small and medium-sized segments are with similar absolute size, as compared to the ground truth.

\begin{figure}[h]
  \centering
  \makebox[0.05\textwidth]{}
  \makebox[0.185\textwidth]{\scriptsize (a)}
  \makebox[0.245\textwidth]{\scriptsize (b)}
  \makebox[0.245\textwidth]{\scriptsize (c)}
  \makebox[0.185\textwidth]{\scriptsize (d)}
  \\
  \raisebox{3.5\height}{\makebox[0.03\textwidth]{\makecell{\scriptsize (1)}}}
  \includegraphics[width=0.23\textwidth]{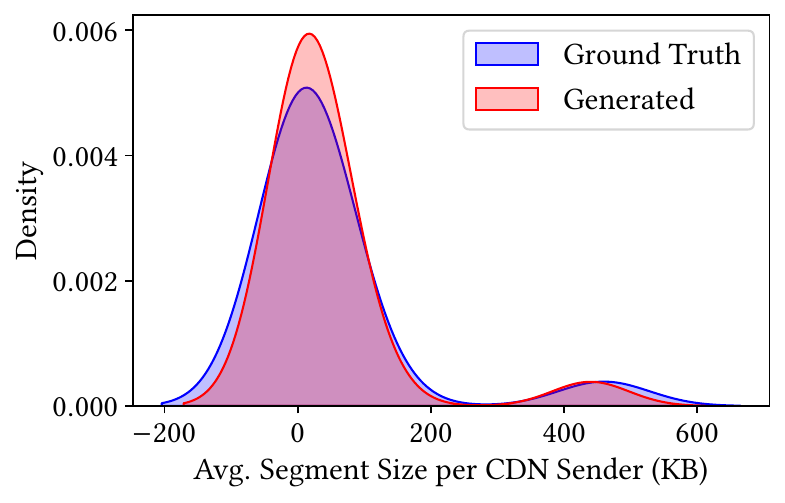}
  \includegraphics[width=0.23\textwidth]{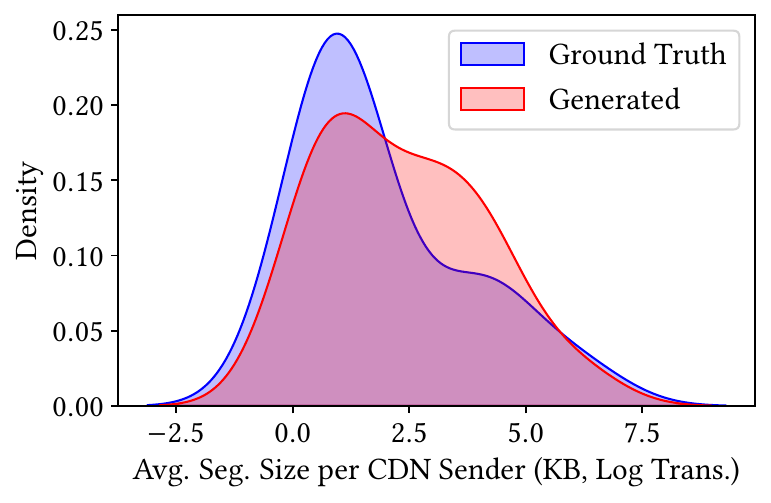}
  \includegraphics[width=0.23\textwidth]{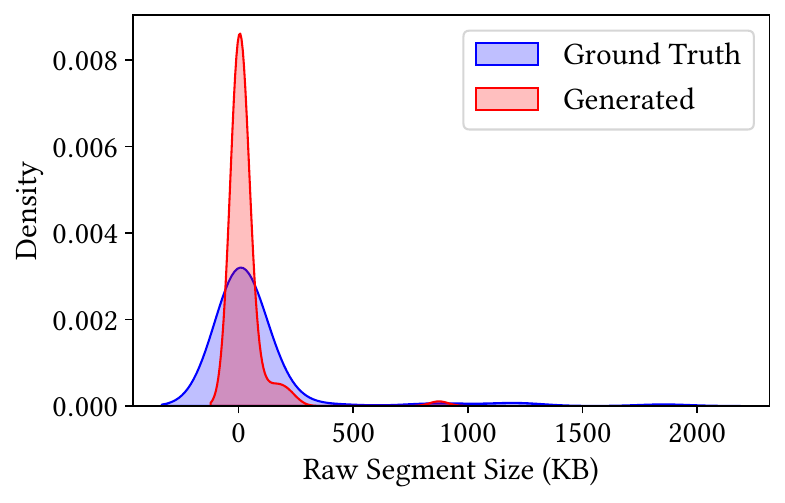}
  \includegraphics[width=0.23\textwidth]{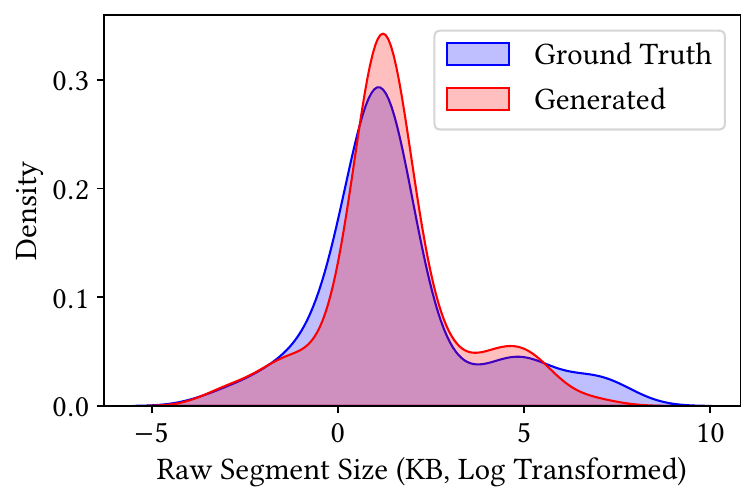}
  \\
  \raisebox{3.5\height}{\makebox[0.03\textwidth]{\makecell{\scriptsize (2)}}}
  \includegraphics[width=0.23\textwidth]{avg_size_kde_30cc5499f5058f639ebb272ebf3fb56a.pdf}
  \includegraphics[width=0.23\textwidth]{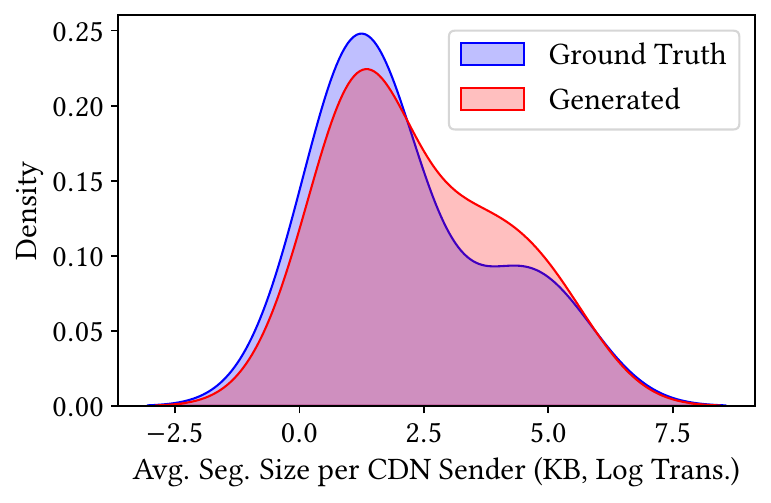}
  \includegraphics[width=0.23\textwidth]{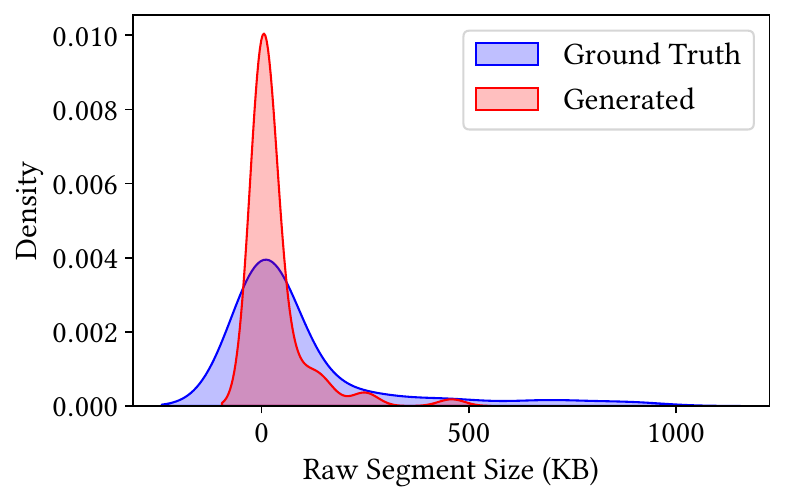}
  \includegraphics[width=0.23\textwidth]{raw_size_log_kde_30cc5499f5058f639ebb272ebf3fb56a.pdf}
  \\
  \raisebox{3.5\height}{\makebox[0.03\textwidth]{\makecell{\scriptsize (3)}}}
  \includegraphics[width=0.23\textwidth]{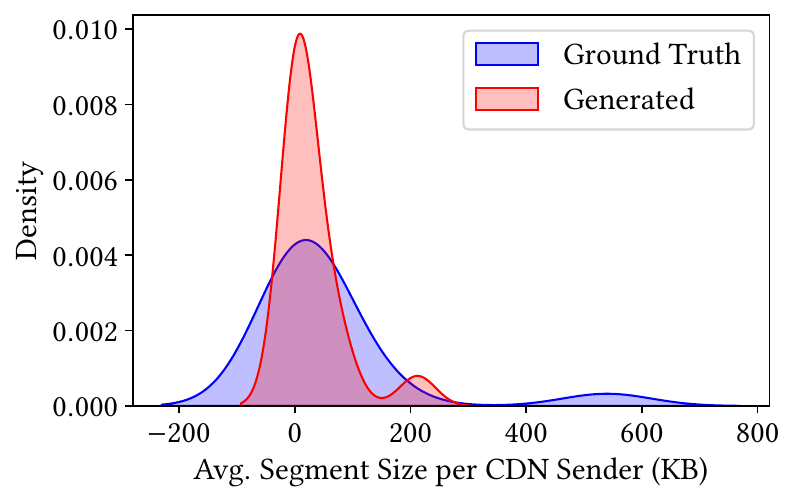}
  \includegraphics[width=0.23\textwidth]{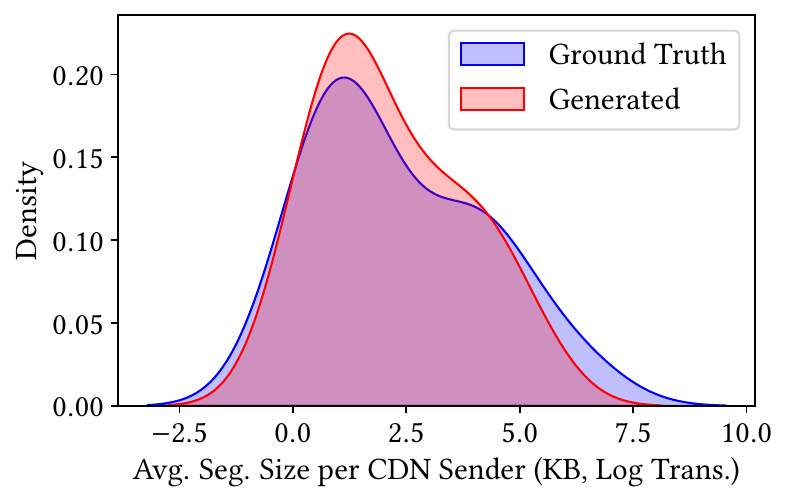}
  \includegraphics[width=0.23\textwidth]{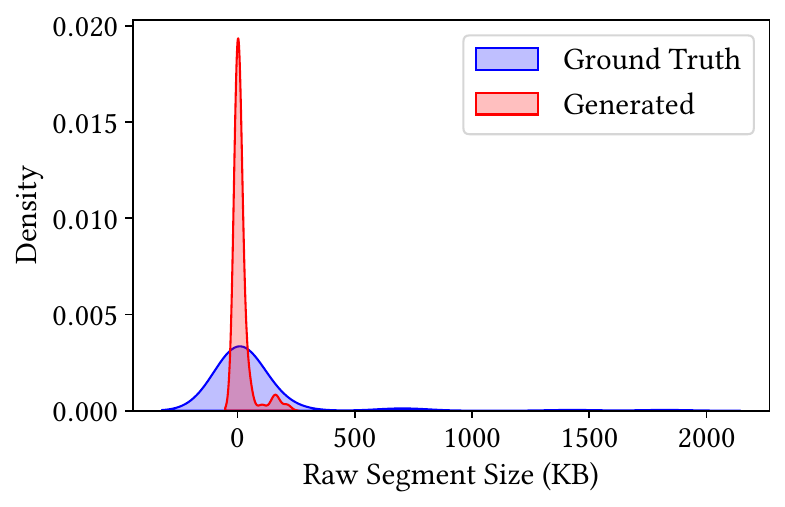}
  \includegraphics[width=0.23\textwidth]{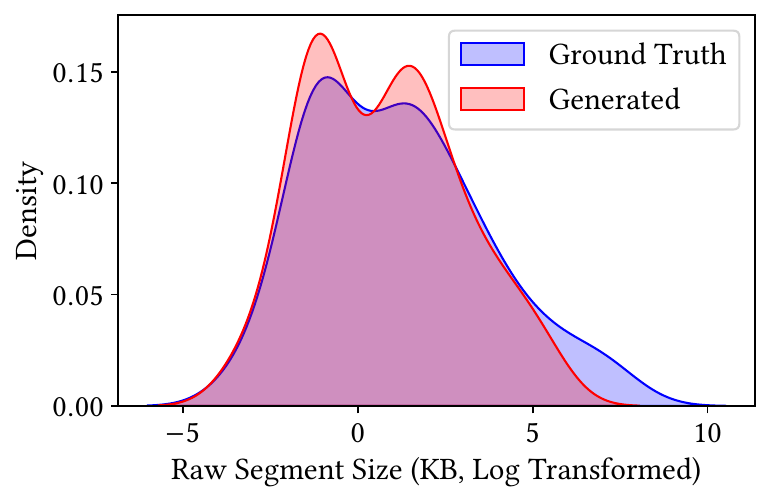}
  \\
  \raisebox{3.5\height}{\makebox[0.03\textwidth]{\makecell{\scriptsize (4)}}}
  \includegraphics[width=0.23\textwidth]{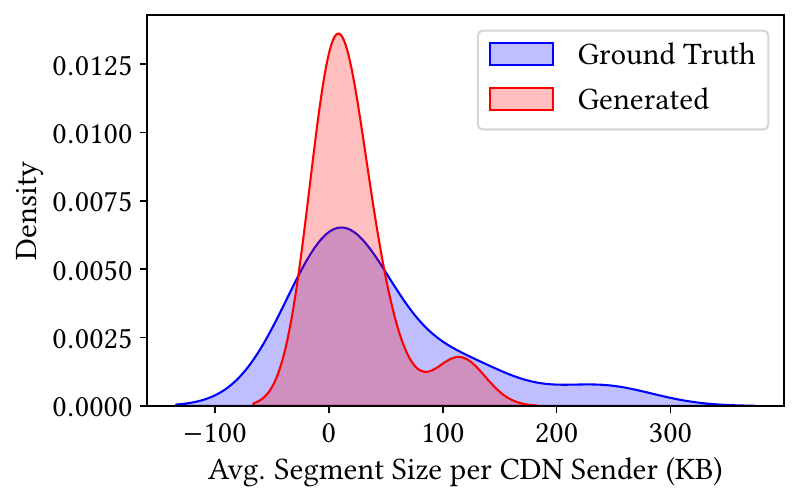}
  \includegraphics[width=0.23\textwidth]{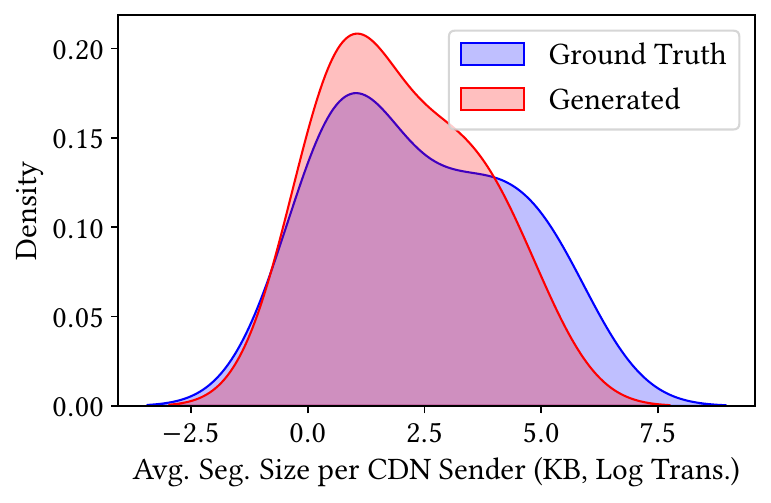}
  \includegraphics[width=0.23\textwidth]{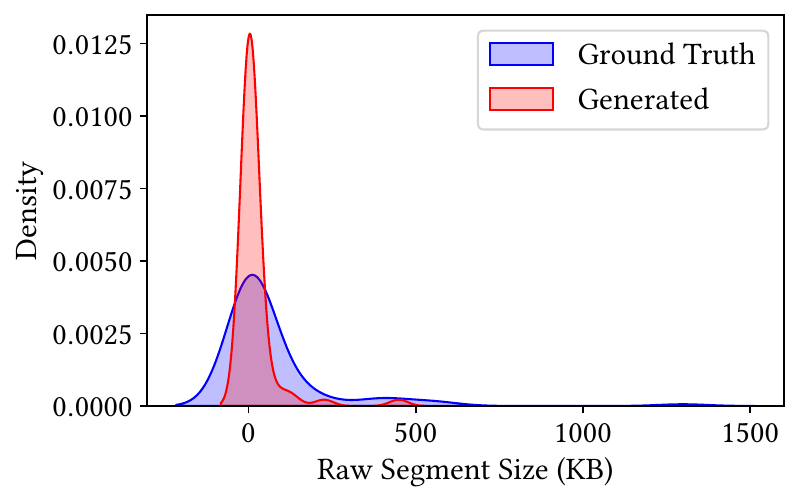}
  \includegraphics[width=0.23\textwidth]{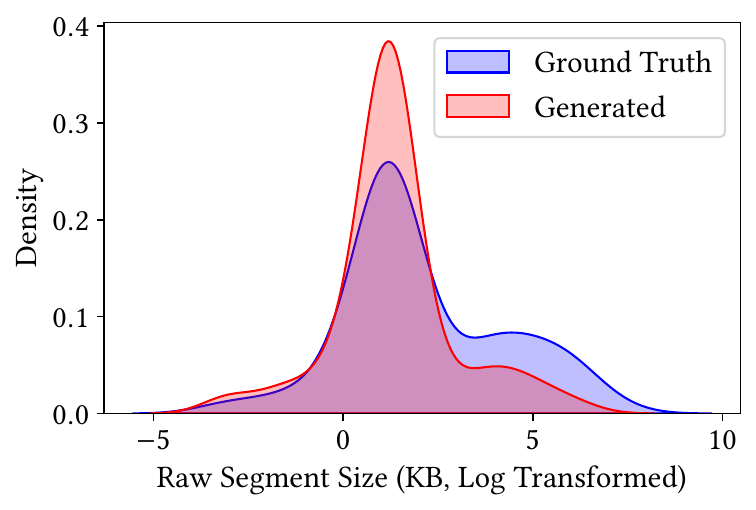}
  \\
  \raisebox{3.5\height}{\makebox[0.03\textwidth]{\makecell{\scriptsize (5)}}}
  \includegraphics[width=0.23\textwidth]{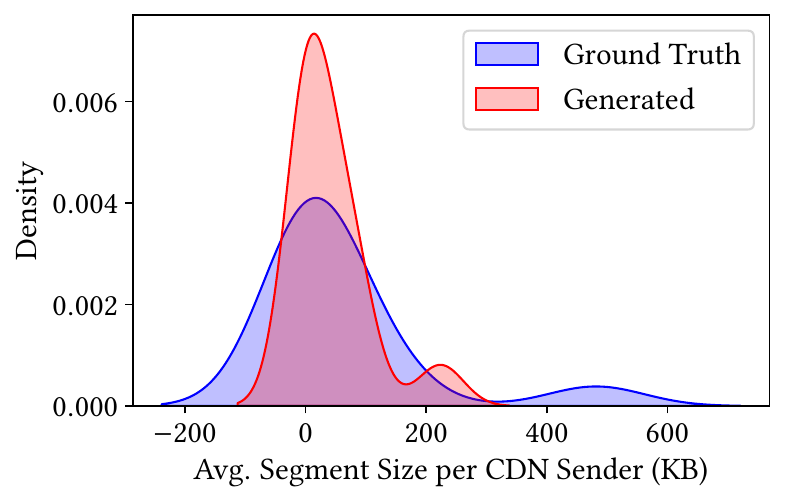}
  \includegraphics[width=0.23\textwidth]{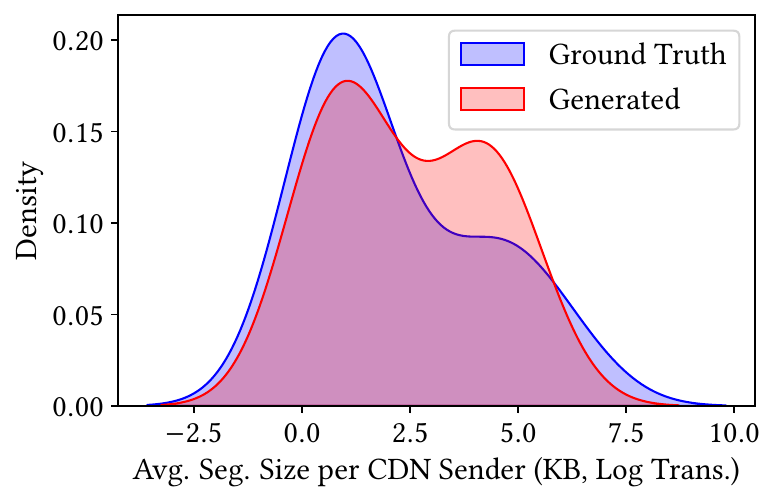}
  \includegraphics[width=0.23\textwidth]{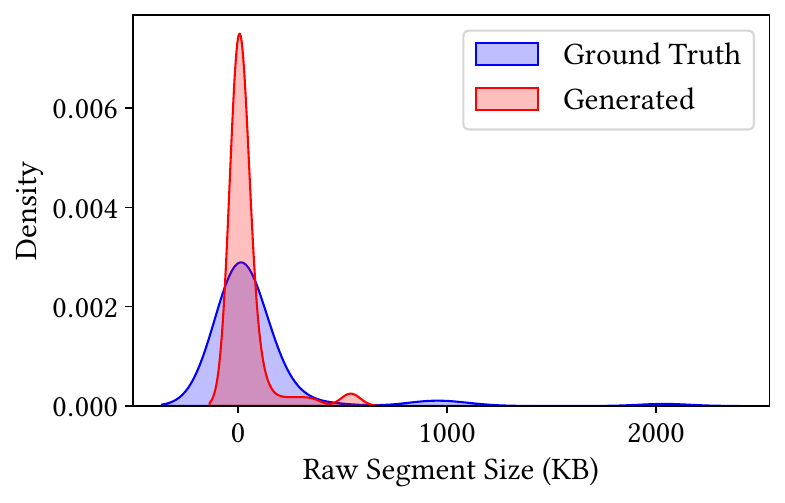}
  \includegraphics[width=0.23\textwidth]{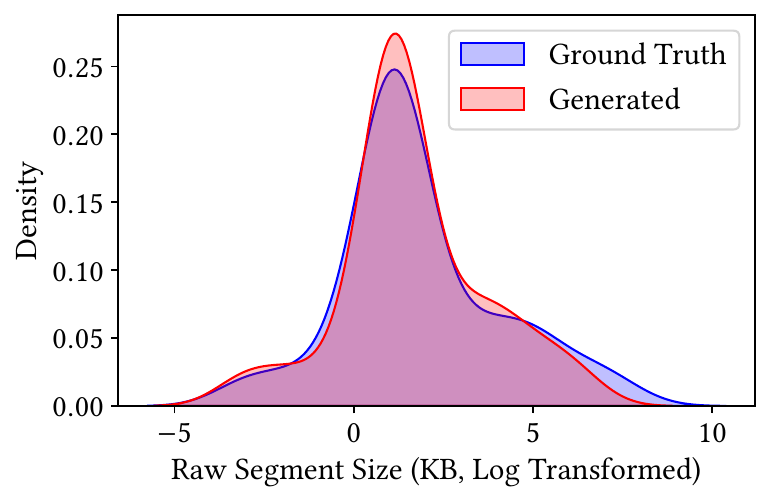}
  \\
  \raisebox{3.5\height}{\makebox[0.03\textwidth]{\makecell{\scriptsize (6)}}}
  \includegraphics[width=0.23\textwidth]{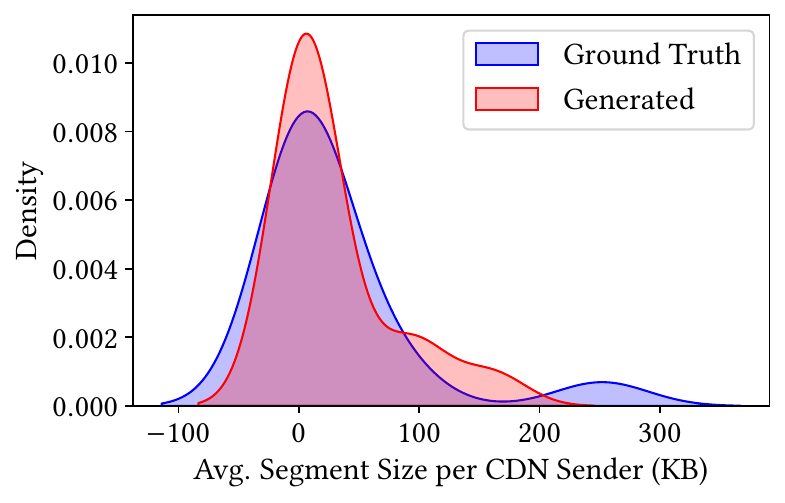}
  \includegraphics[width=0.23\textwidth]{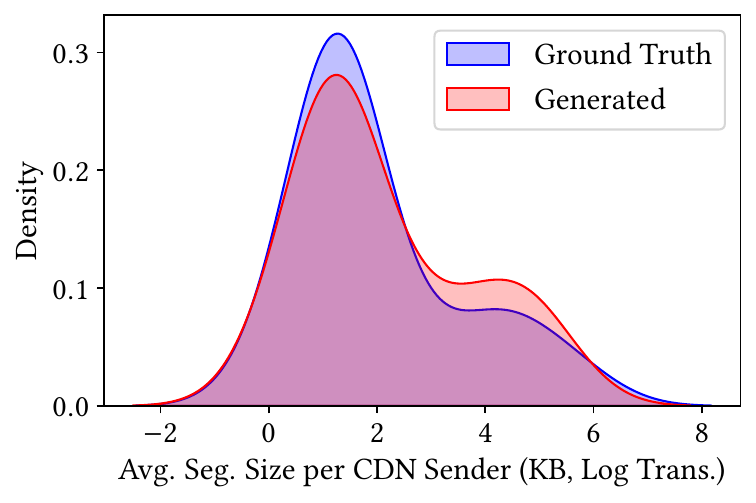}
  \includegraphics[width=0.23\textwidth]{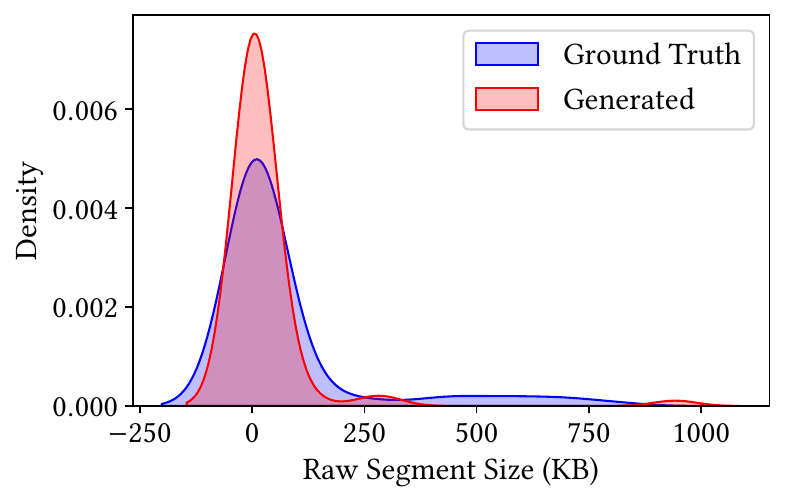}
  \includegraphics[width=0.23\textwidth]{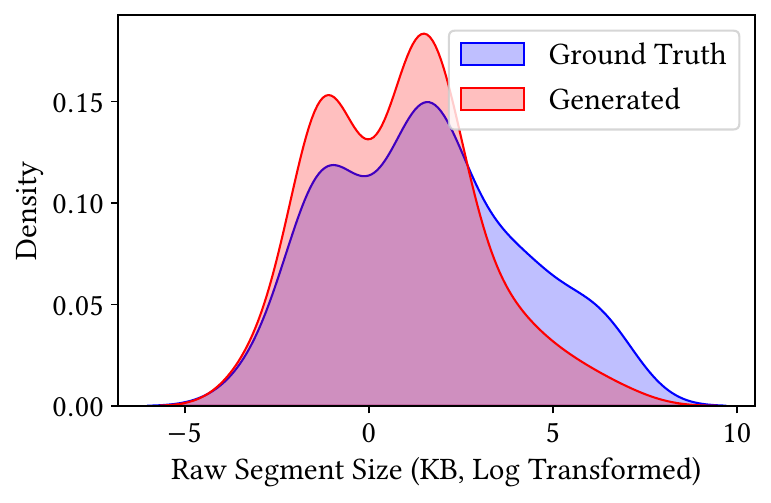}
  \\
  \vspace{-3mm}
  \caption{KDE plots for downloaded segment sizes.}
  \label{fig:full_segment_sizes}
\end{figure}

Figure~\ref{fig:full_segment_sizes} shows additional visualizations for both the average downloaded segment sizes and raw downloaded segments sizes.
Specifically: (a) KDE plots for the average downloaded segment sizes per sender, (b) KDE plots for the log-transformed average downloaded segment sizes per sender, (c) KDE plots for the sizes of all downloaded segments and (d) KDE plots for the log-transformed sizes of all downloaded segments.

\noindent\textbf{Number of Downloaded Segments.}
We also evaluate the number of segments downloaded both in \system's synthetic traces and in the ground truth.
Similar to evaluation of segments' sizes, we find that \system produces data that closely aligns with the ground truth traffic.
In Figures~\ref{fig:segment_analysis_c} and \ref{fig:segment_analysis_e}, there again exists clear overlap in the KDE plots for number of segments downloaded between the ground truth and synthetic data, though it appears \system's traces may not completely capture the tail end cases of higher volume senders.
The quartile values from the Figure~\ref{fig:segment_analysis_e} ECDF further supports the overlap, with only small deltas between the ground truth $(7.00, 10.00, 12.75)$ and generated $(5.75, 9.50, 11.75)$ number of segments downloaded, respectively.

\begin{figure}[t]
  \centering
  \makebox[0.05\textwidth]{}
  \makebox[0.47\textwidth]{\system-generated Netflix}
  \makebox[0.43\textwidth]{Ground truth Netflix}
  \\
  \raisebox{5.5\height}{\makebox[0.03\textwidth]{\makecell{\scriptsize (1)}}}
  \includegraphics[width=0.46\textwidth]{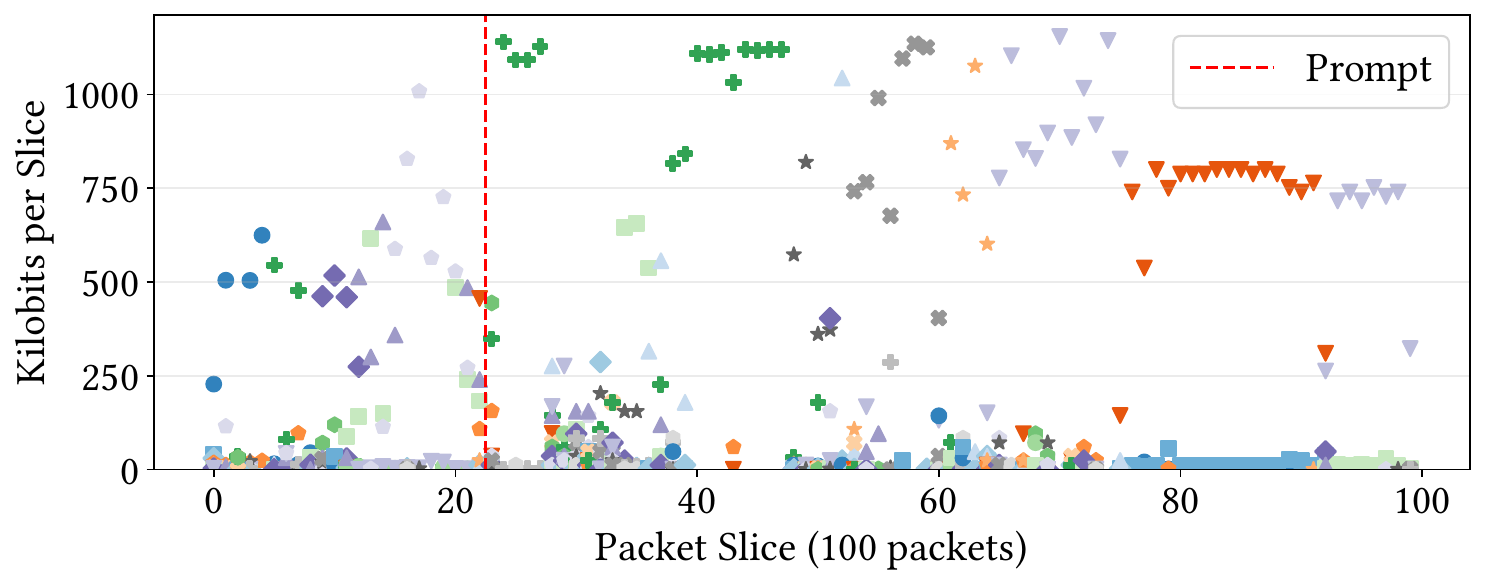}
  \includegraphics[width=0.46\textwidth]{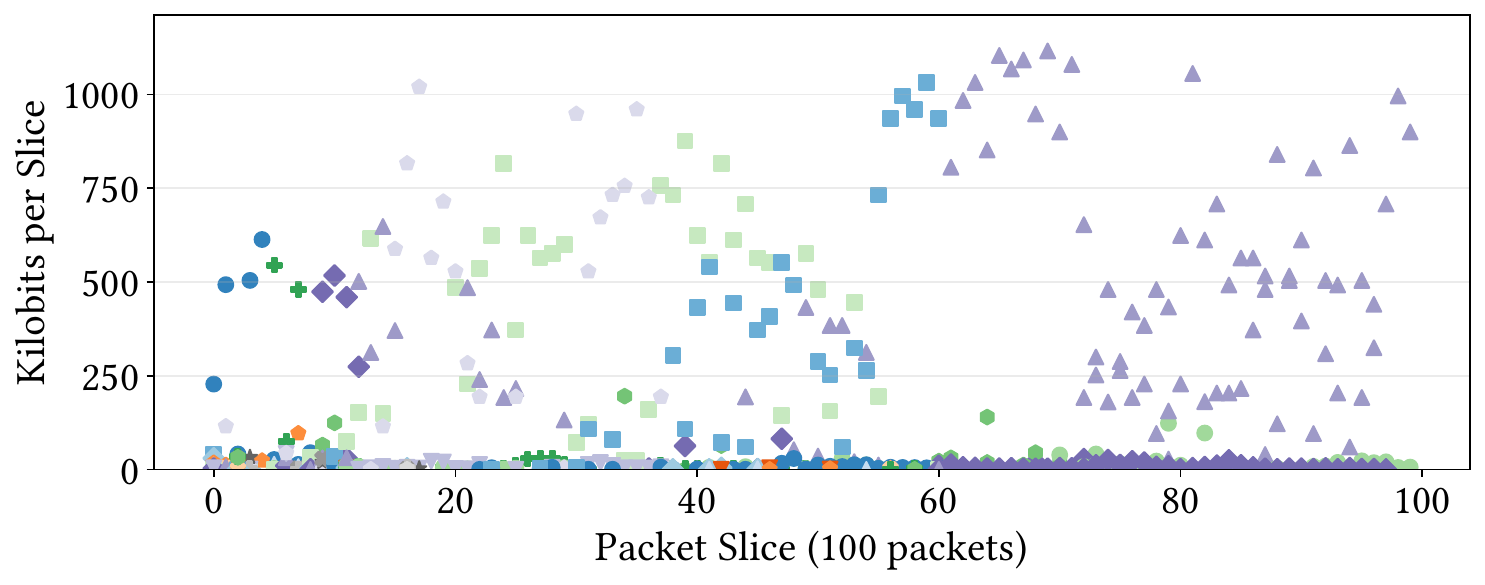}
  \\
  \raisebox{5.5\height}{\makebox[0.03\textwidth]{\makecell{\scriptsize (2)}}}
  \includegraphics[width=0.46\textwidth]{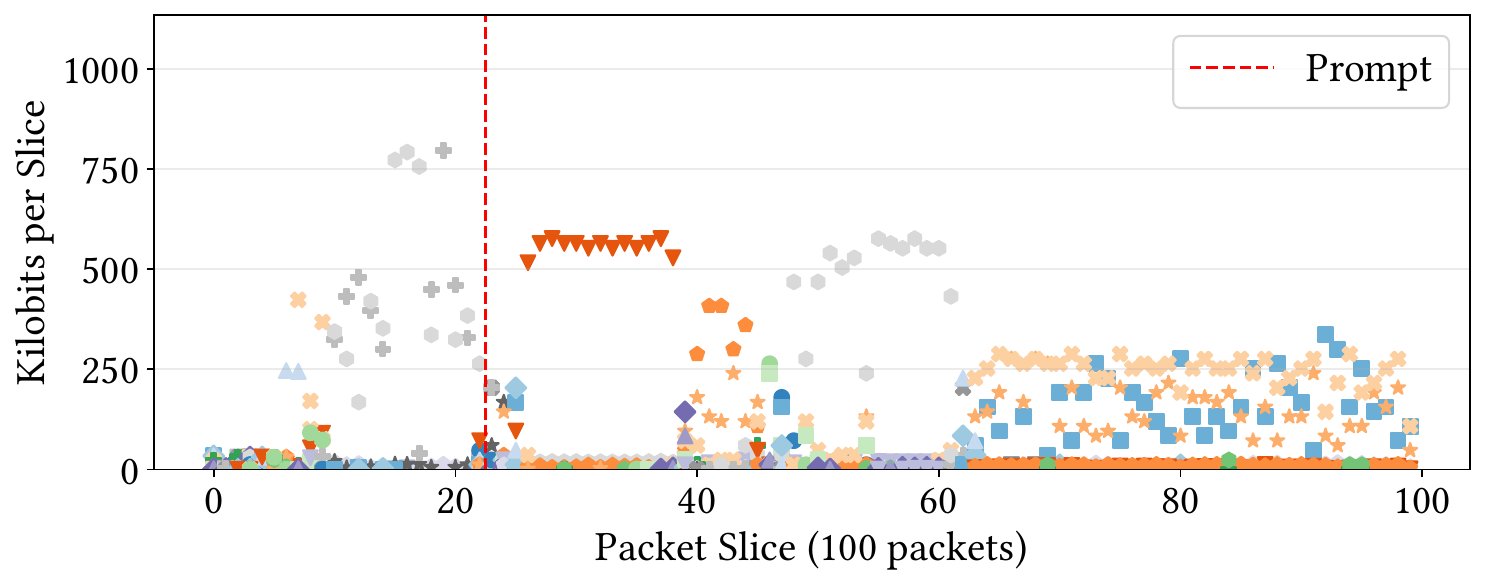}
  \includegraphics[width=0.46\textwidth]{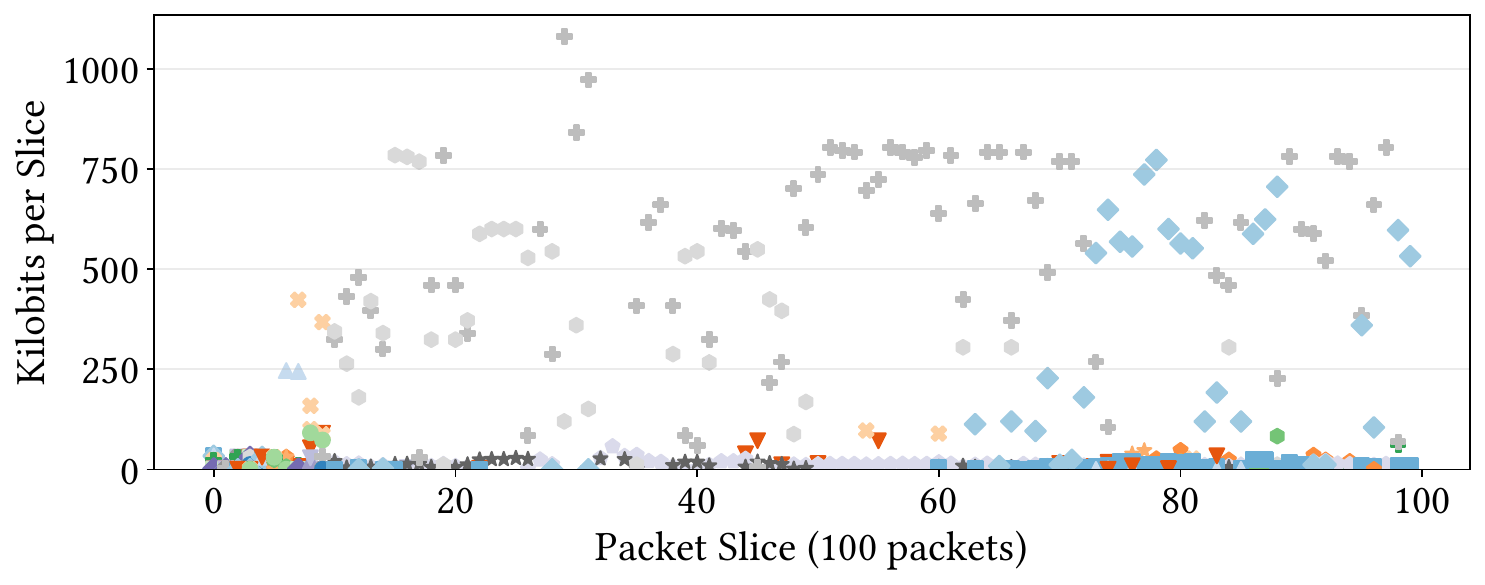}
  \\
  \raisebox{5.5\height}{\makebox[0.03\textwidth]{\makecell{\scriptsize (3)}}}
  \includegraphics[width=0.46\textwidth]{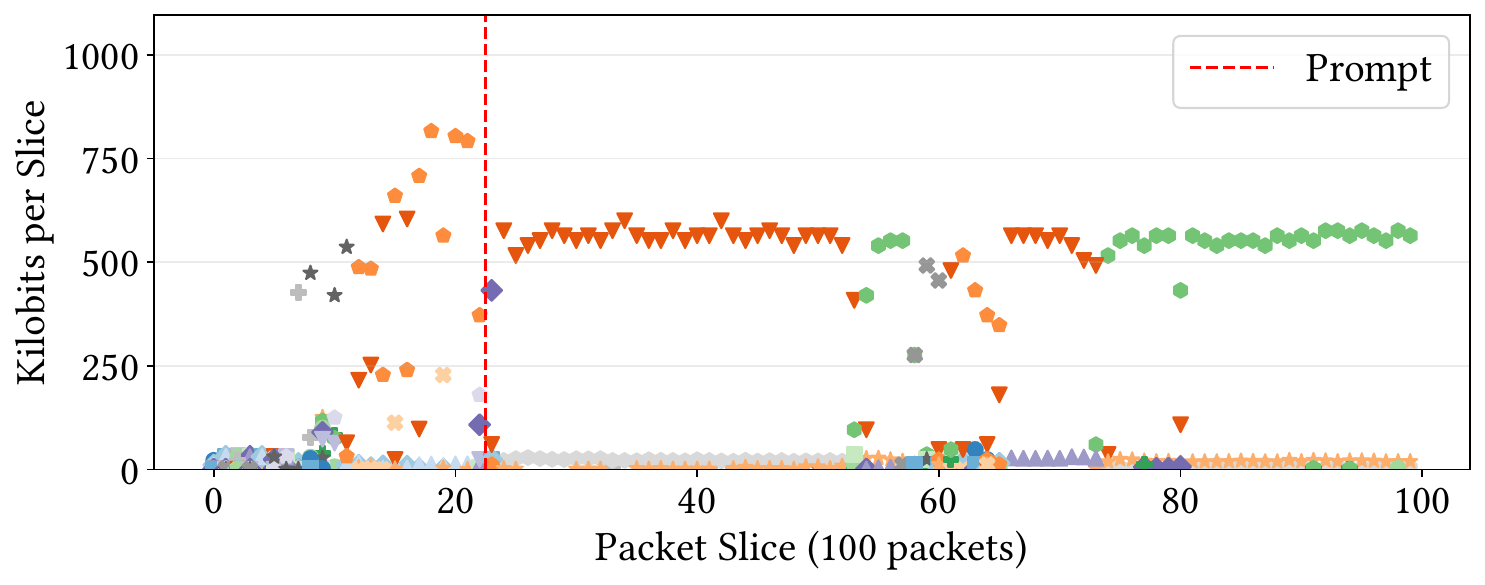}
  \includegraphics[width=0.45\textwidth]{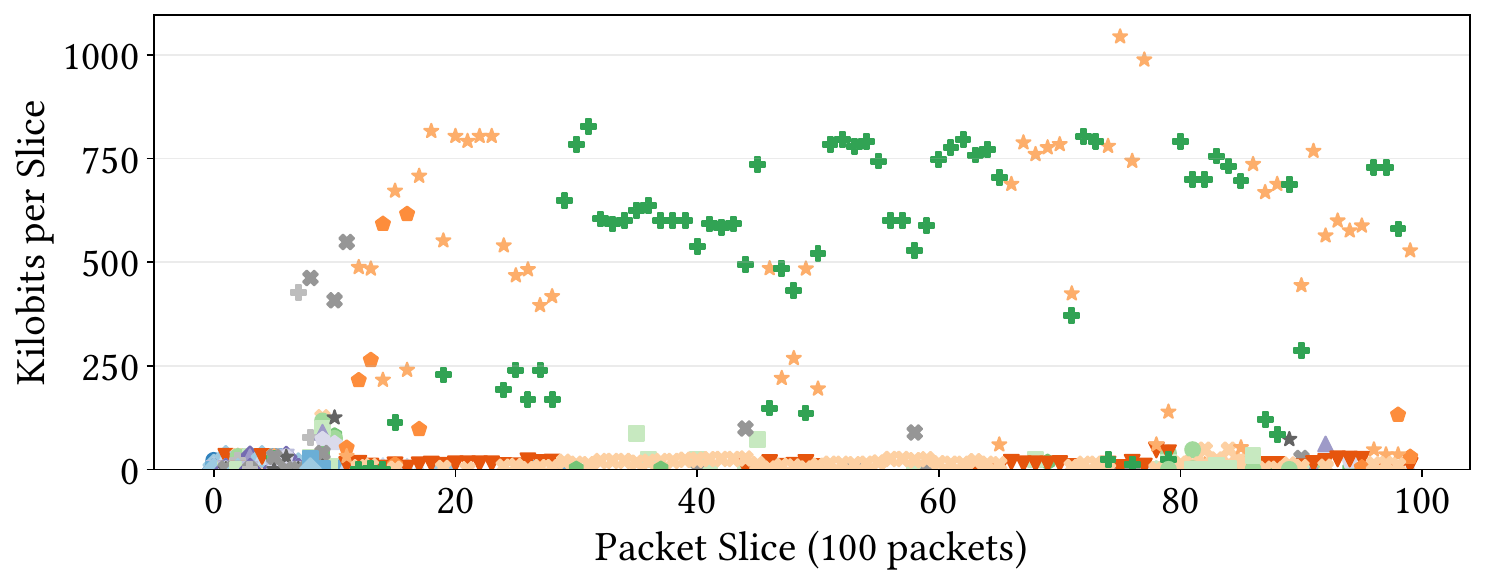}
  \\
  \vspace{1.5em}
  \makebox[0.05\textwidth]{}
  \makebox[0.47\textwidth]{\system-generated YouTube}
  \makebox[0.43\textwidth]{Ground truth YouTube}
  \\
  \raisebox{5.5\height}{\makebox[0.03\textwidth]{\makecell{\scriptsize (1)}}}
  \includegraphics[width=0.46\textwidth]{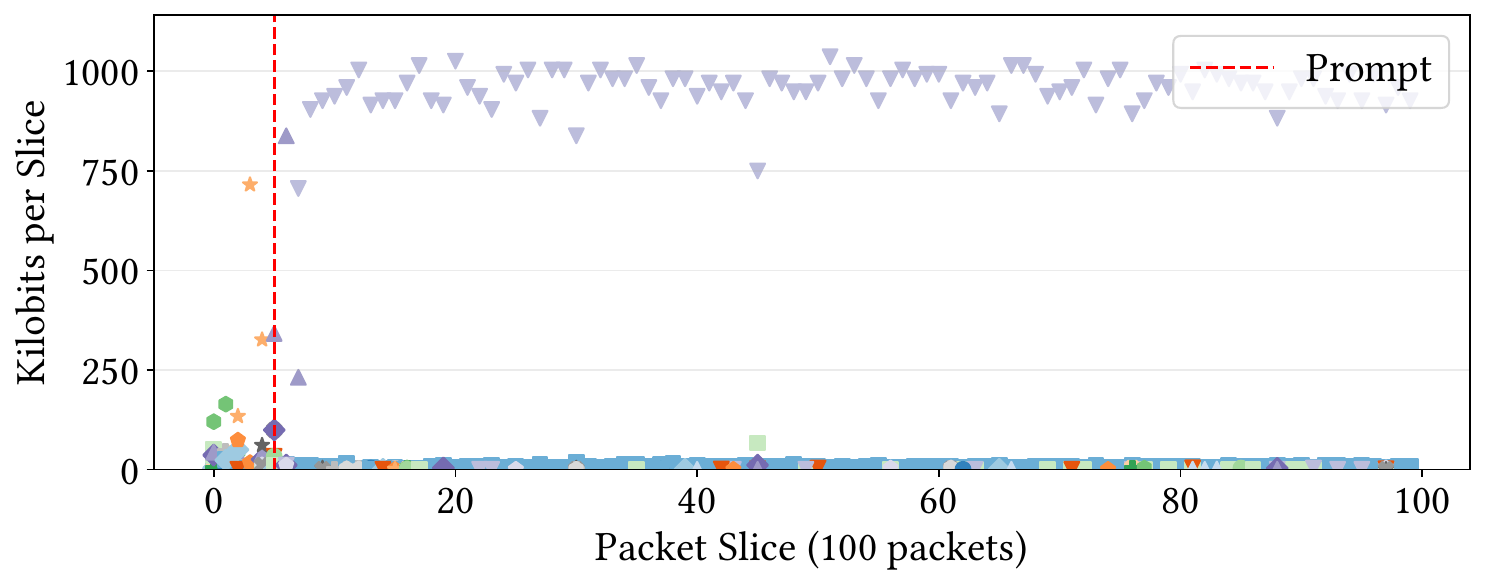}
  \includegraphics[width=0.46\textwidth]{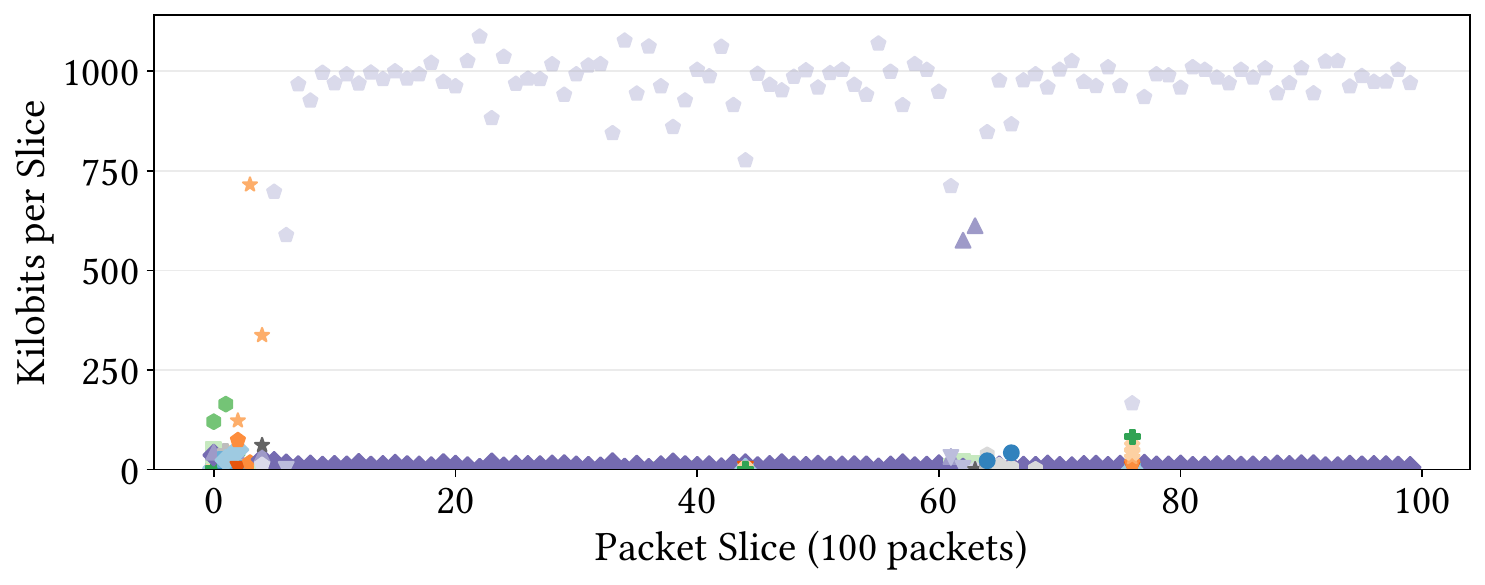}
  \\
  \raisebox{5.5\height}{\makebox[0.03\textwidth]{\makecell{\scriptsize (2)}}}
  \includegraphics[width=0.46\textwidth]{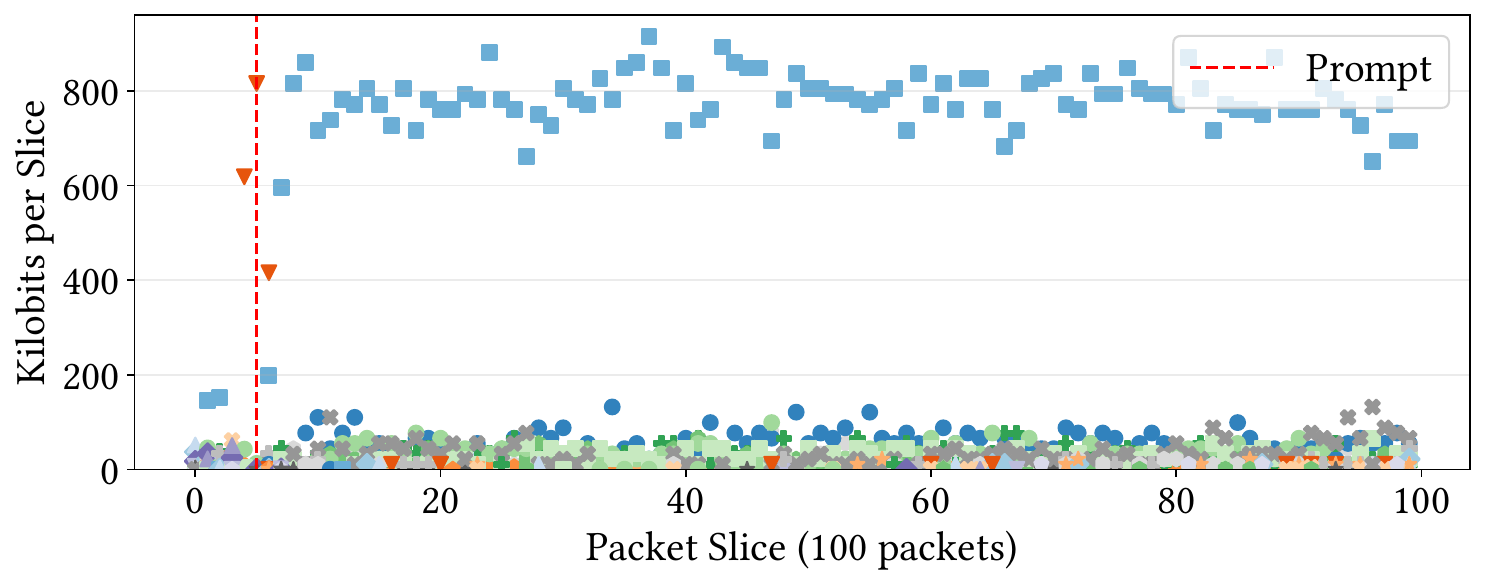}
  \includegraphics[width=0.46\textwidth]{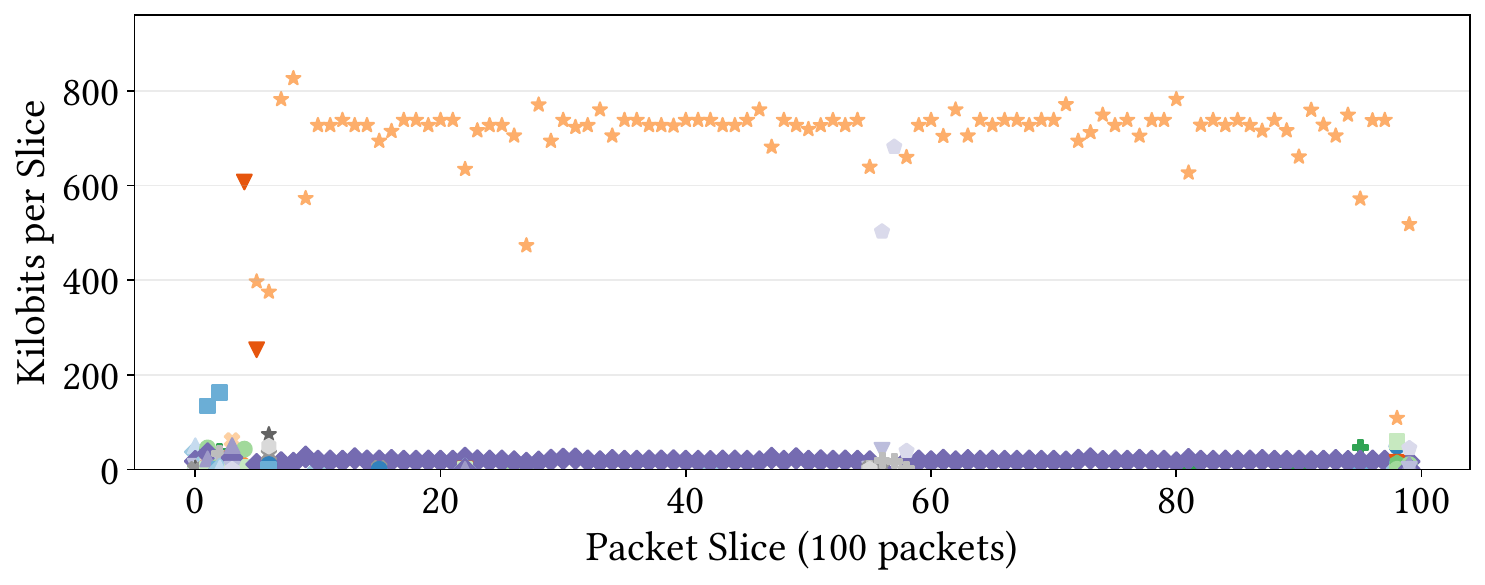}
  \\
  \raisebox{5.5\height}{\makebox[0.03\textwidth]{\makecell{\scriptsize (3)}}}
  \includegraphics[width=0.46\textwidth]{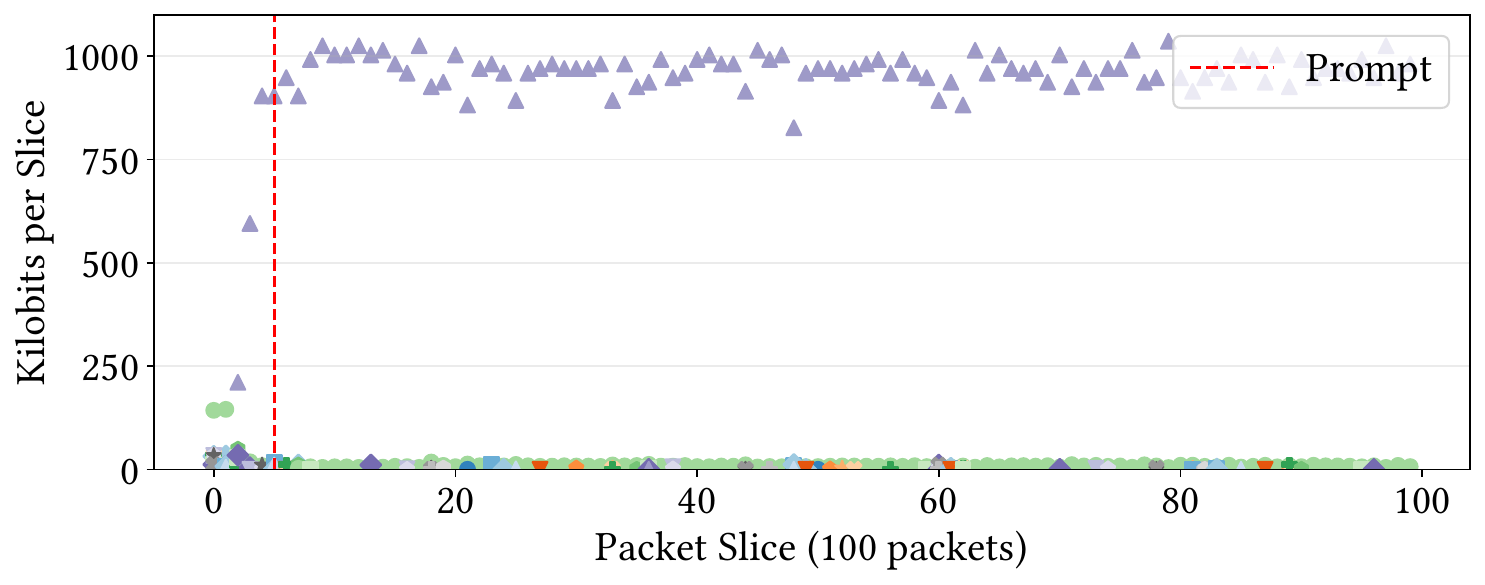}
  \includegraphics[width=0.46\textwidth]{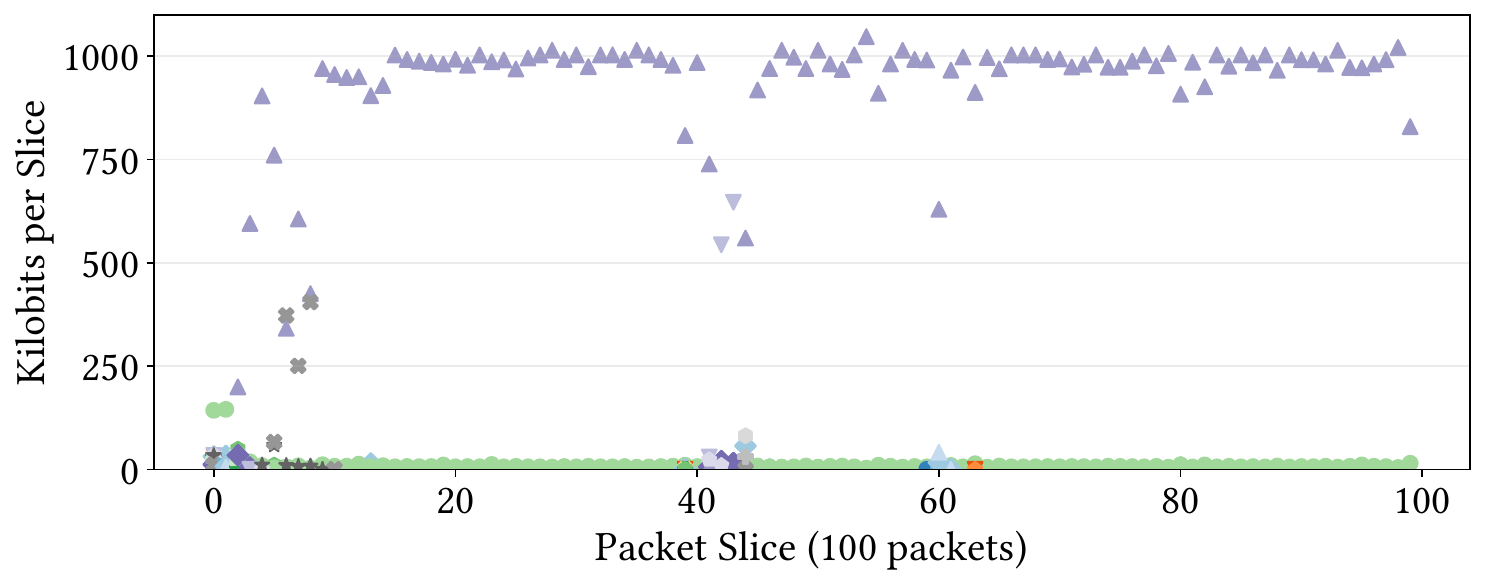}
  \\
  \caption{Plots of throughput per flow for \system-generated/ground truth Netflix and YouTube trace pairs.}
  \label{fig:throughput_extra}
\end{figure}

\newpage
\section{Semantic Similarity}\label{sec:gen_params}

Figure $4$ presents additional examples plotting the throughput in kilobits/100 packet slice for Netflix and YouTube traces.
Table~\ref{tab:gen_params} shows the generation hyperparameters used in Section~\ref{subsec:semantic_sim}.

\begin{table}[!h]
  \centering
  \caption{Generation parameters for \system single and multi-flow models. RP: repetition penalty; T: temperature; MP: min-p; TK: top-k; TP: top-p.}
  \label{tab:gen_params}
  \begin{tabular}{lccccc}
    \toprule
    \textbf{Dataset} & \textbf{RP} & \textbf{T} & \textbf{MP} & \textbf{TK} & \textbf{TP} \\
    \midrule
    Netflix          & 1.8         & 0.15       & 0           & 25          & 0.9         \\
    YouTube          & 1.8         & 0.75       & 0           & 25          & 0.9         \\
    \bottomrule
  \end{tabular}
\end{table}

\section{Memorization Analysis}

% Table~\ref{tab:mem_overall} presents the full results for Section~\ref{subsec:memorization}.

\begin{table*}[!h]
  \centering
  \caption{\sysname{}
    Memorization Analysis Overview.
    Table~\ref{tab:mem_a} reports packet-level memorization and diversity metrics, while
  Table~\ref{tab:mem_b} lists header fields with the largest real–synthetic changes.}
  \label{tab:mem_overall}

  % ============================================================
  % Subtable (a)
  % ============================================================
  \begin{subtable}[t]{0.63\textwidth}
    \centering
    \caption{Packet-Level Memorization and Diversity Metrics}
    \label{tab:mem_a}
    \small

    \begin{tabular}{l r r}
      \toprule
      \multicolumn{3}{c}{\textbf{Basic Comparison}}                     \\
      \midrule
      \sc{Metric}                         & \sc{Value}        & \sc{\%} \\
      \midrule
      Identical Packets                   & --                & 2.35    \\
      Differing Bytes                     & --                & 22.27   \\
      Avg diff. per Packet                & 44.78 bytes       & --      \\
      \midrule

      \multicolumn{3}{c}{\textbf{Intra-Set Diversity}}                  \\
      \midrule
      \textit{Synthetic Packets:}                                       \\
      \quad Avg Pairwise Dist             & 0.359 (norm.)     & --      \\
      \quad Std.
      Dev                                 & 0.206 (norm.)
      & --                          \\
      \textit{Real Packets:}                                            \\
      \quad Avg Pairwise Dist             & 0.680 (norm.)     & --      \\
      \quad Std.
      Dev                                 & 0.250 (norm.)
      & --                          \\
      \textbf{Diversity Ratio (Syn/Real)} & 0.528 (norm.)     & --      \\
      \midrule

      \multicolumn{3}{c}{\textbf{Nearest-Neighbor Memorization}}        \\
      \midrule
      Overall Mean Dist                   & 0.186 (norm.)     & --      \\
      Median Dist                         & 0.170 (norm.)     & --      \\
      Std.
      Dev                                 & 0.085 (norm.)
      & --                          \\
      Min / Max Dist                      & 0.000 / 0.543     & --      \\
      \textit{Thresholds:}                                              \\
      \quad Within 5\%                    & --                & 3.83    \\
      \quad Within 10\%                   & --                & 10.67   \\
      \quad Within 15\%                   & --                & 40.48   \\
      \quad Within 20\%                   & --                & 66.82   \\
      \midrule

      \multicolumn{3}{c}{\textbf{Position-Aware Memorization}}          \\
      \midrule
      \textit{Packets 0--10:}                                           \\
      \quad Avg Dist                      & 0.128 $\pm$ 0.041 & --      \\
      \textit{Packets 10--50:}                                          \\
      \quad Avg Dist                      & 0.175 $\pm$ 0.052 & --      \\
      \textit{Packets 50--100:}                                         \\
      \quad Avg Dist                      & 0.223 $\pm$ 0.083 & --      \\
      \bottomrule
    \end{tabular}
  \end{subtable}
  \hfill
  % ============================================================
  % Subtable (b)
  % ============================================================
  \begin{subtable}[t]{0.34\textwidth}
    \centering
    \caption{Header Fields with Largest Avg Change}
    \label{tab:mem_b}
    \small

    \begin{tabular}{l r}
      \toprule
      Header Field  & Avg Change (bytes) \\
      \midrule
      TCP\_ack      & 835.2M             \\
      TCP\_seq      & 758.8M             \\
      TCP\_chksum   & 19{,}885           \\
      TCP\_sport    & 12{,}974           \\
      TCP\_dport    & 12{,}971           \\
      IP\_id        & 12{,}122           \\
      IP\_chksum    & 11{,}436           \\
      TCP\_window   & 2{,}543            \\
      IP\_len       & 192                \\
      TCP\_urgptr   & 24.68              \\
      IP\_ttl       & 11.97              \\
      IP\_tos       & 5.54               \\
      Raw\_load     & 1.00               \\
      Padding\_load & 1.00               \\
      TCP\_options  & 0.83               \\
      \bottomrule
    \end{tabular}
  \end{subtable}

\end{table*}

\end{document}